\documentclass[12pt]{article}
\usepackage{dcolumn} 
\usepackage{bm} 
\usepackage{a4wide}
\usepackage{fullpage}
\usepackage{slashbox}
\usepackage{tabularx} 
\usepackage{epsf}
\usepackage{amsmath,amsfonts,amssymb}
\usepackage[dvips]{graphicx}
\usepackage{array}
\newcounter{definition}

\def\theequation{\thesection.\arabic{equation}}
\def\appendix{
  \setcounter{section}{0}
  \setcounter{subsection}{0}
  \par
  \def\thesection{Appendix \Alph{section}}
  \def\theequation{\Alph{section}.\arabic{equation}}
  \def\thefigure{\Alph{section}.\arabic{figure}}
}
\newcommand{\nc}{\newcommand}
\nc{\rnc}{\renewcommand}
\rnc{\arraystretch}{1.3}
\nc{\nn}{\nonumber}
\nc{\db}{\displaybreak[0]\\}
\rnc{\d}{\Delta}
\nc{\lam}{\lambda}
\nc{\kp}{\kappa}
\nc{\sg}{\sigma}
\nc{\Pk}[1]{P^{\kp}_{#1}}
\nc{\Ak}[1]{A^{\kp}_{#1}}
\nc{\Qk}[1]{Q^{\kp}_{#1}}
\nc{\lams}[2]{\lam_{#1},\ldots,\lam_{#2}}
\nc{\Mlams}[2]{-\lam_{#1},\ldots,-\lam_{#2}}
\nc{\lamsB}[1]{\lam_1,\lam_2|\ldots|\lam_{2l-1},\lam_{2l}||\lam_{2l+1},\ldots,\lam_{#1}}
\nc{\bra}{\langle}
\nc{\ket}{\rangle}
\nc{\vac}{|0\ket}
\nc{\sh}{\sinh}
\nc{\ch}{\cosh}
\rnc{\(}{\left(}
\rnc{\)}{\right)}
\rnc{\[}{\left[}
\rnc{\]}{\right]}
\nc{\ep}[1]{\epsilon_{#1}}
\nc{\epD}[1]{\epsilon'_{#1}}
\nc{\Tep}[1]{\tilde{\epsilon}_{#1}}
\nc{\TepD}[1]{\tilde{\epsilon}'_{#1}}
\nc{\eps}[2]{\ep{#1},\ldots,\ep{#2}}
\nc{\epDs}[2]{\epD{#1},\ldots,\epD{#2}}
\nc{\Meps}[2]{-\ep{#1},\ldots,-\ep{#2}}
\nc{\MepDs}[2]{-\epD{#1},\ldots,-\epD{#2}}
\nc{\za}[1]{\zeta_a(#1)}
\nc{\zcor}[2]{\bra S_{#1}^z S_{#2}^z \ket}
\nc{\zcorf}[8]{\bra S_{#1}^{#5} S_{#2}^{#6} S_{#3}^{#7} S_{#4}^{#8} \ket}
\nc{\zcors}[6]{\bra S_{1}^{#1} S_{2}^{#2} S_{3}^{#3} S_{4}^{#4} S_{5}^{#5} S_{6}^{#6} \ket}
\nc{\vch}[2]
{\left\bra\(\vec{S}_1\times\vec{S}_2\)\cdot\(\vec{S}_{#1}\times\vec{S}_{#2}\)\right\ket}
\nc{\green}[1]{\bra c_1^{\dagger}c_{#1}\ket_f}
\nc{\dimer}[2]{\left\bra\(\vec{S}_1\cdot\vec{S}_2\)
\(\vec{S}_{#1}\cdot\vec{S}_{#2}\)\right\ket}
\title{Exact evaluation of density matrix elements for the Heisenberg chain}
\author{
  Jun Sato           ${}^1$ \thanks{junji@issp.u-tokyo.ac.jp},   \ \
  Masahiro Shiroishi ${}^1$ \thanks{siroisi@issp.u-tokyo.ac.jp}, \ \ 
  Minoru Takahashi   ${}^1$ \thanks{mtaka@issp.u-tokyo.ac.jp}
\\
\\ 
\it ${}^1$
Institute for Solid State Physics, University of Tokyo,\\\it 
  Kashiwanoha 5-1-5, Kashiwa, Chiba 277-8581, Japan\\\it
}
\begin{document}
\maketitle
%
\begin{abstract}
We have obtained all the density matrix elements on six lattice sites for the spin-1/2 Heisenberg chain via the algebraic method based on the quantum Knizhnik-Zamolodchikov equations. 
Several interesting correlation functions, such as chiral correlation functions, dimer-dimer correlation functions, etc... have been analytically evaluated. 
Furthermore we have calculated all the eigenvalues of the density matrix and analyze the eigenvalue-distribution. 
As a result the exact von Neumann entropy for the reduced density matrix on six lattice sites has been obtained. 
\end{abstract}
%
\newpage
\noindent
\section{Introduction}
The spin-1/2 antiferromagnetic Heisenberg XXZ chain is one of the most
fundamental models for one-dimensional quantum magnetism,
which is given by the Hamiltonian
\begin{align}
\mathcal{H}=\sum_{j=1}^{N} 
\( S_{j}^x S_{j+1}^x + S_{j}^y S_{j+1}^y + \d S_{j}^z S_{j+1}^z \),
\label{Ham}
\end{align}
where $S_j^{\alpha} = \sg_j^{\alpha}/2$ with $\sg_j^{\alpha}$ being the 
Pauli matrices acting on the $j$-th site, 
$\d$ is the anisotropy parameter, and $N$ is the number of lattice sites of this system. 
Here we impose the periodic boundary condition $\vec{S}_{j+N}=\vec{S}_j$. 
For $\d>1$, it is called the massive XXZ model where the system is gapful. 
Meanwhile for $-1<\d\leq1$ case, the system is gapless and  called the massless XXZ model. 
Especially we call it XXX model for the isotropic case $\d=1$. 

The exact eigenvalues and eigenvectors of this model can be obtained 
by the Bethe Ansatz method \cite{Bethe, TakaBook}. 
We shall give a brief survey of this method below. 
First we assume the wave function $|\Psi\ket$ 
with $M$ down spins in the form
\begin{align}
|\Psi\ket&=\sum_{1\leq n_1<\cdots<n_M\leq N} \,\,\, \sum_{\sg\in S_n}
A(\sg)\exp\[i\sum_{j=1}^M k_{\sg(j)}n_j\] |n_1,...,n_M\ket, \nn\\
&A(\sg)=\epsilon(\sg)\prod_{l<j}\(e^{i(k_{\sg(l)}+k_{\sg(j)})}
+1-2\d e^{ik_{\sg(l)}}\),
\label{wf}
\end{align}
where $\sg$ denotes an element of the symmetric group $S_n$, 
$\epsilon(\sg)$ is the sign of the permutation,
and $|n_1,...,n_M\ket$ 
signifies the state where the spins at the positions $n_1,\cdots,n_M$ are downward 
and all the other spins directing upward. 
If the quasi-momenta $\{k_1,k_2,\cdots,k_M\}$ satisfy the Bethe ansatz equations
\begin{align}
e^{ik_jN}=\prod_{l\neq j}^M\(-\frac{e^{i(k_j+k_l)}+1-2\d 
e^{ik_j}}{e^{i(k_j+k_l)}+1-2\d e^{ik_l}}\), \qquad (j=1,2,\cdots,M),
\end{align}
the wave function (\ref{wf}) becomes the eigenvector of the Hamiltonian (\ref{Ham}): 
\begin{align}
\mathcal{H}|\Psi\ket=E|\Psi\ket, \qquad E=\frac{N\d}4+\sum_{j=1}^M(\cos k_j-\d).
\end{align}
In the thermodynamic limit $N\to\infty$, the ground state energy per site $e_0$ 
of the massless XXZ model, for example, is computable 
by analyzing the Bethe ansatz equations 
\begin{align}
e_0=\frac{\d}4-\frac{\sin(\pi\nu)}{2\pi}\int^{\infty}_{-\infty}dt \,
\frac{\sh[(1-\nu)t]}{\sh t\ch(\nu t)}, 
\label{e0}
\end{align}
where we parametrize the anisotropy parameter as $\d=\cos(\pi\nu)$. 
Especially in the XXX case $\d=1$, the ground state energy per site (\ref{e0}) 
can be simplified into 
\begin{align}
e_0=\frac14-\ln2.
\label{e0xxx}
\end{align}
In the same way other physical quantities in the thermodynamic limit such as specific heat, 
magnetic susceptibility, elementary excitations, etc..., can be exactly evaluated 
even at finite temperature by the Bethe ansatz method \cite{TakaBook}. 

The exact calculation of the correlation functions, however, is still a difficult problem 
even in the simplest case for static correlation functions at zero temperature. 
The exceptional case is ${\d=0}$, where the system reduces to a lattice free-fermion model 
by the Jordan-Wigner transformation. 
In this case, we can calculate arbitrary correlation functions 
by means of Wick's theorem \cite{Lieb61,McCoy68}. 
However, there have been rapid developments recently 
in the exact evaluations of correlation functions for ${\d\ne0}$ case also.
Below we shall give the historical review of them, concentrating mainly on 
the static correlation functions of the XXX model $\d=1$ at zero temperature 
for the infinite system $N\to\infty$. 
Until quite recently only the correlation functions 
within three lattice sites had been known: 
\begin{align}
\zcor{1}{2}&=\frac{1}{12}-\frac{1}{3}\za{1}=-0.1477157268533151 \cdots , 
\label{first_neighbor} \\
\zcor{1}{3}&=\frac{1}{12}-\frac{4}{3}\za{1}+\za{3}=0.06067976995643530 \cdots , 
\label{second_neighbor}
\end{align}
where $\zeta_a(s)$ is the alternating zeta function defined by 
${\zeta_a(s) \equiv \sum_{n=1}^{\infty} (-1)^{n-1}/n^s}$, 
which is related to Riemann zeta function ${\zeta(s) \equiv \sum_{n=1}^{\infty} 1/n^s}$ 
as ${\zeta_a(s) = (1-2^{1-s})\zeta(s)}$. 
Note that the alternating zeta function is regular at ${s=1}$ 
and is given by ${\zeta_a(1)}=\ln 2$. 
The nearest-neighbor correlation function (\ref{first_neighbor}) is derived directly from 
the ground state energy (\ref{e0xxx}) obtained by Hulth\'{e}n in 1938 \cite{Hulthen}. 
The first nontrivial correlation function (\ref{second_neighbor})
was derived by Takahashi in 1977 via the strong coupling expansion 
of the ground state energy for the half-filled Hubbard model \cite{Taka77}. 
Note also that another derivation of the second-neighbor correlation function 
(\ref{second_neighbor}) was given by Dittrich and Inozemtsev in 1997 \cite{DI}.
These method, however, can not be generalized to calculate 
further correlation functions on more than four lattice sites, unfortunately. 

Meanwhile using the representation theory of the quantum affine algebra
$U_q(\hat{sl_2})$, Kyoto Group (Jimbo, Miki, Miwa, Nakayashiki) 
derived a multiple integral representation for arbitrary correlation functions of the 
massive XXZ antiferromagnetic chain in 1992 \cite{JMMN, JMBook},
which is before long extended to the XXX case \cite{Nakayashiki, KIEU} 
and the massless XXZ case \cite{JM}. 
More recently the same integral representations were reproduced by 
Kitanine, Maillet, Terras \cite{KMT} in the framework of 
Quantum Inverse Scattering Method. 
They have also succeeded in generalizing the integral representations to the 
XXZ model with an external magnetic field \cite{KMT}. 
More recently the integral formulas are extended to the XXZ model 
even at finite temperature ! \cite{GKS04,GKS05}. 
In this way it has been established now the correlation functions on $n$ lattice sites are 
represented by $n$-dimensional integrals in general. 
However, these multiple integral representations for correlation functions, 
though completely exact, have not been used widely especially among physicists. 
It is mainly because we can not evaluate accurate numerical values of correlation functions 
directly from the integral representation. 
Also it had been a puzzle that the exact expression of the correlation function 
on three lattice sites (\ref{second_neighbor}) had not been reproduced 
from the integral representation for a long time. 

The situation changed when Boos and Korepin devised an innovative method 
to evaluate these multiple integrals for XXX chain in 2001 \cite{BK1, BK2}. 
They showed that the integrand in the multiple integral formula can be reduced to a 
{\it canonical} form, which allows us to implement the integration.
This method was at first applied to a special correlation function called Emptiness 
Formation Probability (EFP) \cite{KIEU} which signifies the probability to find 
a ferromagnetic string of length $n$: 
\begin{align}
P(n) \equiv \left\bra \prod_{j=1}^{n} \( \frac{1}{2} +S_j^z \) \right\ket.
\label{efp}
\end{align}
By performing the multiple integrals, 
the explicit forms of the EFP for up to $n=5$ was obtained \cite{BK1, BK2, BKNS}. 

The Boos-Korepin method was applied 
to calculate other correlation functions on four lattice sites 
in 2003 \cite{SSNT}. 
Especially the third-neighbor correlation function was obtained there as 
\begin{align}
\zcor{1}{4}=&
\frac{1}{12}-3\za{1}+\frac{74}{9}\za{3}-\frac{56}{9}\za{1}\za{3}
-\frac{8}{3}\za{3}^2-\frac{50}{9}\za{5}+\frac{80}{9}\za{1}\za{5}\nn\\
=&-0.05024862725723524\cdots.
\end{align}
Other correlation functions on four lattice sites are given in Appendix A. 

Subsequently, the Boos-Korepin method was generalized to XXZ chain both in massless and 
massive regime and all the correlation functions within four lattice sites were obtained 
for general anisotropy \cite{KSTS03, TKS, KSTS04}. 

In principle, multiple integrals for any correlation functions 
can be performed by means of Boos-Korepin method. 
However, $P(5)$ is the only correlation function which was calculated by this method 
on five lattice sites, since it is getting critically harder to reduce 
the integrand to canonical form as the integral dimension increases. 

In the course of attempting to obtain further correlation functions,
the alternative algebraic method to calculate the EFP was developed by 
Boos, Korepin and Smirnov in 2003 \cite{BKS}. 
They considered the inhomogeneous XXX model, 
in which each site carries an inhomogeneous parameter $\lam_j$. 
The homogeneous XXX model corresponds to the case with 
all the inhomogeneous parameters $\lam_j$ set to be $0$.
Inhomogeneous correlation functions on $n$ lattice sites are 
considered to be functions of variables $\lams{1}{n}$. 
They derived the functional relations which the inhomogeneous EFP should satisfy
by investigating the quantum Knizhnik-Zamolodchikov (qKZ) equations 
\cite{KZ, FR, Smirnov1, Smirnov2}, 
the solutions to which are connected with 
the inhomogeneous correlation functions \cite{JMBook, Nakayashiki, JM}. 
Moreover they suggested an ansatz for the form of the inhomogeneous EFP, 
which consists of only one transcendental function $\omega(\lam)$ 
with rational functions of inhomogeneous parameters $\lams{1}{n}$ as coefficients. 
(for the proof of the ansatz and further generalizations, 
see \cite{BJMST1, BJMST2, BJMST3, BJMST4, BJMST5, BJMST6}). 
It was shown that the functional relations together with the ansatz for the final form 
completely fix the explicit form of the inhomogeneous EFP. 
In this way the explicit forms of the inhomogeneous EFP for up to $n=6$ have been obtained, 
which gives a new result for $P(6)$ in the homogeneous limit $\lam_j\to0$. 

This Boos-Korepin-Smirnov method based on the qKZ equation was generalized to 
arbitrary correlation functions in \cite{BST}. 
Actually, in that paper, all the correlation functions on five lattice sites are obtained 
based on the functional relations for the general correlation functions (see Appendix A). 
Especially the fourth-neighbor correlation function is given by 
\begin{align}
\zcor{1}{5}=&
\frac{1}{12}-\frac{16}{3}\za{1}+\frac{290}{9}\za{3}-72\za{1}\za{3}-\frac{1172}{9}\za{3}^2
-\frac{700}{9}\za{5} \nn\\& +\frac{4640}{9}\za{1}\za{5}-\frac{220}{9}\za{3}\za{5}
-\frac{400}{3}\za{5}^2+\frac{455}{9}\za{7}-\frac{3920}{9}\za{1}\za{7} \nn\\& 
+280\za{3}\za{7} = 0.03465277698272816 \cdots. 
\end{align}
The main purpose of this paper is to report further results 
using the algebraic method in \cite{BST}. 
More explicitly we have succeeded in calculating all the correlation functions 
on six lattice sites using this algebraic method. 
We remark that, if we consider only the two-point correlation functions $\zcor{1}{n}$, 
it was already calculated up to the seventh-neighbor correlation function 
$\zcor{1}{8}$ \cite{SS, SST}. 
The method actually allows us to evaluate some other correlation functions 
such as string correlation functions \cite{BSS}, but not all the correlation functions. 
We also remark that there is a related but different approach for the evaluation of 
the general correlation functions, 
which is developed recently by Boos, Jimbo, Miwa, Smirnov and Takeyama 
\cite{BJMST1, BJMST2, BJMST3, BJMST4, BJMST5, BJMST6}. 
They have established a compact exponential formula for general correlation functions 
without heavy multiple integrals, which would be also useful 
to evaluate correlation functions analytically. 
We, however, do not discuss the formula in this paper. 

The plan of this paper is as follows. 
In section 2, the algebraic method to calculate general correlation functions is reviewed. 
In section 3, we present the analytical results for some physically interesting correlation 
functions, such as chiral correlation functions, dimer-dimer correlation functions, 
etc$\cdots$. 
In section 4, we show the eigenvalue-distribution of the reduced density matrix 
and calculate the von Neumann entropy. 
Summary and discussion are given in section 5. 
\section{Algebraic method to evaluate general correlation functions 
based on the qKZ equations}
\setcounter{equation}{0}
Below we describe the functional approach to evaluate general correlation functions 
established in \cite{BST}. 
Any correlation function can be expressed as a sum of 
density matrix elements $P^{\epDs{1}{n}}_{\eps{1}{n}}$, 
which are defined by the ground state expectation value 
of the product of elementary matrices: 
\begin{align}
P^{\epDs{1}{n}}_{\eps{1}{n}}\equiv\bra E_1^{\epD{1}\ep{1}} \cdots E_n^{\epD{n}\ep{n}} \ket, 
\end{align}
where
$E_j^{\epD{j}\ep{j}}$ are $2 \times 2$ elementary matrices acting on the $j$-th site as 
\begin{align}
E^{++}_j&=\begin{pmatrix}1&0\\0&0\\\end{pmatrix}_{\!\![j]}=\frac12+S_j^z, \quad
E^{--}_j=\begin{pmatrix}0&0\\0&1\\\end{pmatrix}_{\!\![j]}=\frac12-S_j^z, \nn\\
E^{+-}_j&=\begin{pmatrix}0&1\\0&0\\\end{pmatrix}_{\!\![j]}=S_j^+=S_j^x+i S_j^y, \quad
E^{-+}_j=\begin{pmatrix}0&0\\1&0\\\end{pmatrix}_{\!\![j]}=S_j^-=S_j^x-i S_j^y. \nn
\end{align}
For example, the density matrix elements on four lattice sites $P^{-++-}_{++--}$ 
can be written in terms of spin operators as 
\begin{align}
P^{-++-}_{++--}=\left\bra S_1^-\(\frac12+S_2^z\)S_3^+\(\frac12-S_4^z\)\right\ket. 
\end{align}
First we rewrite the functional relations for inhomogeneous density matrix elements, 
which follow from the qKZ equations: 
\begin{itemize}
\item{Translational invariance}
\begin{align}
P^{\epDs{1}{n}}_{\eps{1}{n}}(\lam_1 + \lam, \ldots, \lam_n + \lam)=
P^{\epDs{1}{n}}_{\eps{1}{n}}(\lams{1}{n}),
\label{f1}
\end{align}
\item{Transposition, Negating and Reverse order relations}
\begin{align}
&P^{\epDs{1}{n}}_{\eps{1}{n}}(\lams{1}{n})=
P^{\eps{1}{n}}_{\epDs{1}{n}}(\Mlams{1}{n}) \nn\\
&=P^{\MepDs{1}{n}}_{\Meps{1}{n}}(\lams{1}{n})
=P^{\epDs{n}{1}}_{\eps{n}{1}}(\Mlams{n}{1})
\end{align}
\item{Intertwining relation}
\begin{align}
&\sum_{\TepD{j},\TepD{j+1}=\pm}
R^{\epD{j}\epD{j+1}}_{\TepD{j}\TepD{j+1}}(\lam_j-\lam_{j+1})
P^{\ldots,\TepD{j+1},\TepD{j},\ldots}_{\ldots,\ep{j+1},\ep{j},\ldots}
(\ldots, \lam_{j+1},\lam_j,\ldots) \nn\\
& = \sum_{\Tep{j},\Tep{j+1}=\pm}
P^{\ldots,\epD{j},\epD{j+1},\ldots}_{\ldots,\Tep{j},\Tep{j+1},\ldots}
(\ldots,\lam_j ,\lam_{j+1},\ldots)
R^{\Tep{j}\Tep{j+1}}_{\ep{j}\ep{j+1}}(\lam_j-\lam_{j+1}),
\end{align}
where $R$ denotes the $R$-matrix of the XXX model:
\begin{align}
R(\lam)=
\begin{pmatrix}
R^{++}_{++}(\lam) & R^{++}_{+-}(\lam) & R^{++}_{-+}(\lam) & R^{++}_{--}(\lam) \\
R^{+-}_{++}(\lam) & R^{+-}_{+-}(\lam) & R^{+-}_{-+}(\lam) & R^{+-}_{--}(\lam) \\
R^{-+}_{++}(\lam) & R^{-+}_{+-}(\lam) & R^{-+}_{-+}(\lam) & R^{-+}_{--}(\lam) \\
R^{--}_{++}(\lam) & R^{--}_{+-}(\lam) & R^{--}_{-+}(\lam) & R^{--}_{--}(\lam) \\
\end{pmatrix}=
\begin{pmatrix}
1 &           0           &          0            & 0 \\
0 & \frac{\lam}{\lam + 1} & \frac{1}{\lam + 1}    & 0 \\
0 & \frac{1}{\lam + 1}    & \frac{\lam}{\lam + 1} & 0 \\
0 &           0           &          0            & 1 \\
\end{pmatrix}.
\end{align}
\item{First recurrent relation}
\begin{align}
&P^{\epD{1}\epDs{2}{n}}_{\ep{1}\eps{2}{n}}(\lam+1,\lam,\lams{3}{n})
=-\delta_{\ep{1},-\ep{2}}\epD{1}\ep{2}
P^{\epD{2}\epDs{3}{n}}_{-\epD{1}\eps{3}{n}}(\lam,\lams{3}{n}) \nn\\
&P^{\epD{1}\epDs{2}{n}}_{\ep{1}\eps{2}{n}}(\lam-1,\lam,\lams{3}{n})
=-\delta_{\epD{1},-\epD{2}}\ep{1}\epD{2}
P^{-\ep{1}\epDs{3}{n}}_{\ep{2}\eps{3}{n}}(\lam,\lams{3}{n})
\label{rec1}
\end{align}
\item{Second recurrent relation}
\begin{align}
\lim_{\lam_{j} \rightarrow i \infty}
P^{\epDs{1}{j},\ldots,\epD{n}}_{\eps{1}{j},\ldots,\ep{n}}(\lams{1}{j},\ldots,\lam_n)
=\delta_{\ep{j},\epD{j}}\frac{1}{2}
P^{\epDs{1}{j-1},\epDs{j+1}{n}}_{\eps{1}{j-1},\eps{j+1}{n}}(\lams{1}{j-1},\lams{j+1}{n})
\label{rec2}
\end{align}
\item{Identity relation}
\begin{align}
\sum_{\eps{1}{n}}
&P^{\epDs{1}{n}}_{\eps{1}{n}}(\lams{1}{n})
=\sum_{\epDs{1}{n}}P^{\epDs{1}{n}}_{\eps{1}{n}}(\lams{1}{n}) \nn\\
&=P^{+,\ldots,+}_{+,\ldots,+}(\lams{1}{n})=P^{-,\ldots,-}_{-,\ldots,-}(\lams{1}{n})
\end{align}
\item{Reduction relation}
\begin{align}
&P^{+,\epDs{2}{n}}_{+,\eps{2}{n}}(\lam_1,\lams{2}{n})+
P^{-,\epDs{2}{n}}_{-,\eps{2}{n}}(\lam_1,\lams{2}{n})
=P^{\epDs{2}{n}}_{\eps{2}{n}}(\lams{2}{n}) \nn\\
&P^{\epDs{1}{n-1},+}_{\eps{1}{n-1},+}(\lams{1}{n-1},\lam_n)
+P^{\epDs{1}{n-1},-}_{\eps{1}{n-1},-}(\lams{1}{n-1},\lam_n)
=P^{\epDs{1}{n-1}}_{\eps{1}{n-1}}(\lams{1}{n-1})
\label{f2}
\end{align}
\end{itemize}

Additionally it is established that density matrix elements can be written 
in the form of \cite{BKS, BJMST1, BST}
\begin{align}
P^{\epDs{1}{n}}_{\eps{1}{n}}&(\lams{1}{n})=
\(\prod^n_{j=1}\frac{\delta_{\ep{j},\epD{j}}}2\)\nn\\
&+\sum_{l=1}^{[\frac{n}{2}]}\sum_{\sg\in T_{n,l}}
A^{\epDs{1}{n}}_{\eps{1}{n}}(l,\sg| \lam_{\sg(1)},\cdots,\lam_{\sg(n)})
\prod_{j=1}^l\omega(\lam_{\sg(2j-1)}-\lam_{\sg(2j)}), 
\label{form}
\end{align}
where $T_{n,l}$ is a subset of the symmetry group of degree $n$ defined as 
\begin{align}
T_{n,l}=\{\sg\in S_n|&\sg(1)<\sg(3)<\cdots<\sg(2l-1), \nn\\
&\sg(2j-1)<\sg(2j) \quad {\rm for} \quad j=1,2,\cdots,l, \nn\\
&\sg(2l+1)<\sg(2l+2)<\cdots<\sg(n) \}. 
\end{align}
$A^{\epDs{1}{n}}_{\eps{1}{n}}(l,\sg|\lams{1}{n})$ 
are rational functions of inhomogeneous parameters $\lams{1}{n}$ 
with known denominator 
\begin{align}
&A^{\epDs{1}{n}}_{\eps{1}{n}}(l,\sg|\lams{1}{n}) = 
\frac{Q^{\epDs{1}{n}}_{\eps{1}{n}}(l,\sg| \lams{1}{n})}{D_{n,l}(\lams{1}{n})}, \nn\\
&D_{n,l}(\lams{1}{n})=
\frac{\prod_{k=1}^l(\lam_{2k-1}-\lam_{2k})
\prod_{2l+1\leq k<j\leq n}(\lam_{k}-\lam_{j})}
{\prod_{1\leq k<j\leq n}(\lam_{k}-\lam_{j})}. 
\end{align}
$Q^{\epDs{1}{n}}_{\eps{1}{n}}(l,\sg| \lams{1}{n})$ 
are polynomials of known degree in 
inhomogeneous parameters $\lams{1}{n}$ with rational coefficients 
which have at most the same degree in each variable and also at most the same total degree 
as in the denominator. 

A transcendental function $\omega(\lam)$ in (\ref{form}) is defined by
\begin{align}
\omega(\lam)=\frac{1}{2}+\sum_{k=1}^{\infty}(-1)^k \frac{2k(\lam^2-1)}{\lam^2-k^2}
=2\sum^{\infty}_{k=0}\lam^{2k}\{\za{2k-1}-\za{2k+1}\}. 
\end{align}
Here note that the identity $\za{-1}=-3\zeta(-1)=1/4$. 
The following properties of the function $\omega(\lam)$ are needed 
for calculations of the recurrent relations (\ref{rec1})-(\ref{rec2}):
\begin{align}
\omega(i \infty)=0, 
\quad
\omega(\lam \pm 1)=\frac{3}{2}\frac{1}{\lam^2-1}
-\frac{\lam(\lam\pm 2)}{\lam^2-1}\omega(\lam). 
\end{align}

Below we list the explicit forms of all the non-zero elements of 
inhomogeneous density matrices for $n=1,\,2$: 
\begin{align}
&P^+_+(\lam_1)=P^-_-(\lam_1)=\frac12, \nn\\
&P^{++}_{++}(\lam_1,\lam_2)=P^{--}_{--}(\lam_1,\lam_2)
=\frac14+\frac16\omega(\lam_1-\lam_2), \nn\\
&P^{+-}_{+-}(\lam_1,\lam_2)=P^{-+}_{-+}(\lam_1,\lam_2)
=\frac14-\frac16\omega(\lam_1-\lam_2), \nn\\
&P^{+-}_{-+}(\lam_1,\lam_2)=P^{-+}_{+-}(\lam_1,\lam_2)=\frac13\omega(\lam_1-\lam_2). 
\end{align}
It can be easily seen that these satisfy the functional relations (\ref{f1})-(\ref{f2}). 
By taking the homogeneous limit $\lam_1,\lam_2\to0$, 
we obtain density matrix elements for the physically interesting homogeneous case:
\begin{align}
&P^+_+=P^-_-=\frac12, 
&P^{++}_{++}=P^{--}_{--}=\frac13-\frac13\za{1}, \nn\\
&P^{+-}_{+-}=P^{-+}_{-+}=\frac16+\frac13\za{1}, 
&P^{+-}_{-+}=P^{-+}_{+-}=\frac16-\frac23\za{1}. 
\end{align}

In the paper \cite{BST}, it has been argued that the functional relations 
(\ref{f1})-(\ref{f2}) together with the ansatz for the form of 
inhomogeneous correlation functions (\ref{form}) completely determine 
the unknown rational coefficients in the polynomials 
$Q^{\epDs{1}{n}}_{\eps{1}{n}}(l,\sg| \lams{1}{n})$ and therefore the full form 
of the inhomogeneous correlation functions. 
As an example, we give the explicit form of the density matrix element for $n=3$
\begin{align}
&Q^{-++}_{++-}(l\!=\!1,\sg\!=\!(123)|\lam_1,\lam_2,\lam_3)=\frac16-\frac1{12}\lam_{12}, \nn\\
&Q^{-++}_{++-}(l\!=\!1,\sg\!=\!(231)|\lam_1,\lam_2,\lam_3)=\frac16+\frac1{12}\lam_{12}, \nn\\
&Q^{-++}_{++-}(l\!=\!1,\sg\!=\!(132)|\lam_1,\lam_2,\lam_3)
=\frac16+\frac16\lam_{13}\lam_{23}-\frac1{12}(\lam_{13}+\lam_{23}), \nn\\
&P^{-++}_{++-}(\lam_1,\lam_2,\lam_3)=\frac{2-\lam_{12}}{12\lam_{13}\lam_{23}}\omega_{12}+
\frac{2+\lam_{23}}{12\lam_{21}\lam_{31}}\omega_{23}+
+\frac{2+2\lam_{12}\lam_{32}-(\lam_{12}+\lam_{32})}{12\lam_{12}\lam_{32}}\omega_{13},
\end{align}
where we have used the abbreviations $\omega_{jk}=\omega(\lam_{jk})$, 
$\lam_{jk}=\lam_{j}-\lam_{k}$. 
By taking the homogeneous limit $\lam_1,\lam_2,\lam_3\to0$, 
we obtain the homogeneous density matrix element
\begin{align}
P^{-++}_{++-}=\frac1{12}-\frac43\za{1}+\za{3}.
\end{align}
In this way all the correlation functions on five lattice sites 
have been obtained in \cite{BST}. 
In this paper we have obtained further results for {\it six} lattice sites. 
There are 24 independent correlation functions among them, 
explicit forms of which are collected in Appendix B. 

Concerning two-point correlation functions $\zcor{1}{n}$, on the other hand, 
an efficient way to calculate them has been developed \cite{SS, SST}. 
Namely, without evaluating all the density matrix elements, 
a two-point correlation function can be derived from its generating function
\begin{align}
\Pk{n} \equiv \left\bra \prod^n_{j=1} \left\{ \(\frac{1}{2}+S^z_j\) 
+\kp\(\frac{1}{2}-S^z_j\) \right\} \right\ket
\label{gf}
\end{align}
through the relation
\begin{align}
\zcor{1}{n}= \frac{1}{2} \frac{\partial^2}{\partial \kp^2} 
\Big\{ \Pk{n} -2\Pk{n-1} +\Pk{n-2} \Big\} \Bigg|_{\kp=1} - \frac{1}{4}.
\end{align}
Note that the generating function $\Pk{n}$ can be considered as a generalization of the 
Emptiness Formation Probability (\ref{efp}), 
which is reproduced if we set $\kp=0$ in (\ref{gf}). 
From the functional relations for general density matrix elements (\ref{f1})-(\ref{f2}), 
it can be shown that the generating function $\Pk{n}$ satisfies 
the following closed functional relations: 
\begin{itemize}
\item{Translational invariance}
\begin{align}
\Pk{n}(\lam_1 + \lam, \ldots, \lam_n + \lam)=\Pk{n}(\lams{1}{n})
\end{align}
\item{Negating relation}
\begin{align}
\Pk{n}(\Mlams{1}{n})=\Pk{n}(\lams{1}{n})
\end{align}
\item{Symmtry relation}
\begin{align}
\Pk{n}(\lams{1}{n})=\Pk{n}(\lams{\sigma(1)}{\sigma(n)}),
\label{sym}
\end{align} 
where $\sigma$ denotes any element of the symmetry group $S_n$. 
\item{First recurrent relation}
\begin{align}
\Pk{n}(\lams{1}{n-1},\lam_{n-1}\pm 1)=\kp \Pk{n-2}(\lams{1}{n-2})
\end{align}
\item{Second recurrent relation}
\begin{align}
\lim_{\lam_n\to i \infty}\Pk{n}(\lams{1}{n-1},\lam_n)=\frac{1+\kp}{2}\Pk{n-1}(\lams{1}{n-1})
\end{align}
\end{itemize}
One can explicitly calculate the generating functions $\Pk{n}$ recursively with respect to 
$n$ from these functional relations together with the ansatz for the final form
\begin{align}
\Pk{n}(\lams{1}{n})&=
\(\frac{1+\kp}2\)^n
+\sum_{l=1}^{[\frac{n}{2}]}\sum_{\sg\in T_{n,l}}
\Ak{n,l}(\lam_{\sg(1)},\cdots,\lam_{\sg(n)})
\prod_{j=1}^l\omega(\lam_{\sg(2j-1)}-\lam_{\sg(2j)}), \nn\\
&\Ak{n,l}(\lams{1}{n}) = \frac{\Qk{n,l}(\lams{1}{n})}{D_{n,l}(\lams{1}{n})}, \nn\\
\end{align}
where $T_{n,l}$, $D_{n,l}(\lams{1}{n})$ and $\omega(\lam)$ are the same as in (\ref{form}) 
and $\Qk{n,l}(\lams{1}{n})$ are polynomials containing the parameter $\kp$. 
The great advantage of this method is that the polynomials $\Qk{n,l}(\lams{1}{n})$ 
do not depend on the permutation $\sg\in T_{n,l}$ due to the symmetry relation (\ref{sym}), 
unlike the case of general density matrix elements 
$Q^{\epDs{1}{n}}_{\eps{1}{n}}(l,\sg| \lams{1}{n})$. 
This fact considerably reduces the amount of bothersome calculation. 
Actually by using this method, up to the seventh-neighbor correlation function $\zcor{1}{8}$ 
have been calculated \cite{SST}. 
Calculating all the density matrix elements is much harder. 
We, therefore, have succeeded only up to six sites yet. 
\section{Applications to physically interesting correlation functions}
\setcounter{equation}{0}
In this section, we discuss several physically interesting correlation functions,
which can be evaluated exactly from our results of 
all the density matrix elements for $n=6$. 
\subsection{Chiral correlation function}
First let us consider the vector chiral correlation function 
\begin{align}
\vch{j}{j+1}=6\(\zcorf{1}{2}{j}{j+1}{x}{z}{x}{z}-\zcorf{1}{2}{j}{j+1}{x}{z}{z}{x}\),
\end{align}
which measures a chirality of the spin alignment. 
It has been observed that they have simple expressions \cite{BST, MT} 
compared with the other correlation functions on the same lattice sites 
\begin{align}
\vch{3}{4}&=\frac12\(\za{1}-\za{3}\)=-0.1041977484048752\cdots,\\
\vch{4}{5}&=\frac12\(\za{1}-\za{3}\)-\frac54\(\za{3}-\za{5}\)=-0.01597638205835821\cdots.
\end{align}
Here we have newly obtained the vector chiral correlation function for six lattice sites, 
which has also a simple form though it contains some quadratic terms 
\begin{align}
\vch{5}{6}&=\frac12\za{1}-\frac{11}3\za{3}+9\za{5}-\frac{35}6\za{7}+\frac43\za{1}\za{3}\nn\\&
\quad+\frac43\za{3}^2-\frac{32}3\za{1}\za{5}-\frac43\za{3}\za{5}+\frac{28}3\za{1}\za{7}\nn\\
&=-0.01774606473688137\cdots.
\end{align}
It is interesting to note that the numerical values of these are all negative and 
their absolute values are oscillating.

Furthermore we obtain the scalar chiral correlation function defined as below, 
which reveals an intriguing factorization:
\begin{align}
&\left\bra\[\(\vec{S}_1\times\vec{S}_2\)\cdot\vec{S}_3\]
\[\(\vec{S}_4\times\vec{S}_5\)\cdot\vec{S}_6\]\right\ket
\nn\\&\qquad=6(\zcors{x}{y}{z}{x}{y}{z}+\zcors{x}{y}{z}{y}{z}{x}+\zcors{x}{y}{z}{z}{x}{y}
\nn\\&\qquad\qquad-\zcors{x}{y}{z}{z}{y}{x}-2\zcors{x}{y}{z}{x}{z}{y})
\nn\\&\qquad=\frac76(\za{1}-1/4)(\za{7}-\za{5})-\frac16(\za{3}-\za{1})(\za{5}-\za{3})
\nn\\&\qquad=0.008133862120680087\cdots.
\end{align}
Note that the numerical value in this case is positive. 
These numerical values may indicate a classical picture of the spin alignment for 
the antiferromagnetic ground state of quantum Heisenberg chain. 
\subsection{One-particle Green function}
By the Jordan-Wigner transformation
\begin{align}
S_j^+=\prod_{k=1}^{j-1}(1-2c_k^{\dagger}c_k)c_j, \quad
S_j^-=\prod_{k=1}^{j-1}(1-2c_k^{\dagger}c_k)c_j^{\dagger}, 
\end{align}
the XXX model is transformed into the isotropic spinless fermion model. 
Let us consider the one-particle Green function $\green{n}$ for this model. 
Here the bracket $\bra\cdots\ket_f$ means the expectation value in the half-filled 
state of the spinless fermion model. 
The first-neighbor one-particle Green function $\green{2}$ is directly obtained from 
the Hulthen's result (\ref{first_neighbor}) as 
\begin{align}
\green{2}=\bra S_1^+S_2^-\ket=2\zcor{1}{2}=\frac16-\frac23\za{1}=-0.2954314537066302\cdots. 
\end{align}
The first non-trivial result for this one-particle Green function
has been obtained in \cite{SSNT} 
\begin{align}
\green{4}&=4\bra S_1^+S_2^zS_3^zS_4^-\ket=8\zcorf{1}{2}{3}{4}{x}{z}{z}{x}\nn\\
&=\frac1{30}-2\za{1}+\frac{338}{45}\za{3}-\frac{40}{9}\za{1}\za{3}-\frac{32}{15}\za{3}^2-\frac{52}{9}\za{5}+\frac{64}{9}\za{1}\za{5}\nn\\&=0.08228771668643698\cdots. 
\end{align}
because $\green{n}=0$ if $n$ is odd. 
Here we have newly obtained the fifth-neighbor one-particle Green function
\begin{align}
\green{6}&=16\bra S_1^+S_2^zS_3^zS_4^zS_5^zS_6^-\ket=32\zcors{x}{z}{z}{z}{z}{x}\nn\\
&=\frac{1}{70}-\frac{10}{3}\za{1}+74\za{3}-\frac{3608}{9}\za{1}\za{3}-\frac{90832}{45}\za{3}^2-\frac{100288}{135}\za{3}^3\nn\db&-\frac{207464}{315}\za{5}+\frac{133456}{15}\za{1}\za{5}+\frac{3088}{9}\za{3}\za{5}+\frac{200576}{45}\za{1}\za{3}\za{5}\nn\db&-\frac{60704}{45}\za{3}^2\za{5}-\frac{1943840}{63}\za{5}^2+\frac{242816}{9}\za{1}\za{5}^2+\frac{46112}{9}\za{3}\za{5}^2\nn\db&+\frac{490880}{189}\za{5}^3+\frac{89918}{45}\za{7}-\frac{308392}{9}\za{1}\za{7}+\frac{623128}{9}\za{3}\za{7}\nn\db&-\frac{424928}{9}\za{1}\za{3}\za{7}-\frac{645568}{45}\za{3}^2\za{7}+\frac{89536}{9}\za{5}\za{7}\nn\db&-\frac{98176}{9}\za{3}\za{5}\za{7}-\frac{236432}{9}\za{7}^2+\frac{343616}{9}\za{1}\za{7}^2-\frac{7052}{5}\za{9}\nn\db&+\frac{128928}{5}\za{1}\za{9}-\frac{196304}{3}\za{3}\za{9}+\frac{645568}{15}\za{1}\za{3}\za{9}\nn\db&+\frac{98176}{5}\za{3}^2\za{9}+\frac{135104}{3}\za{5}\za{9}-\frac{196352}{3}\za{1}\za{5}\za{9}
\nn\\&=-0.04497471675792834\cdots. 
\end{align}
It will be an interesting problem to study the asymptotic behavior of one-particle Green 
function in detail. 
Conformal field theory predict that
\begin{align}
\green{x}\sim \cos\(k_Fx\)x^{-5/4},\quad k_F=\pi/2, 
\end{align}
omitting the logarithmic correction \cite{KY}. 
Unfortunately our new exact results are not sufficient to confirm  the asymptotic formula. 
\subsection{Dimer-dimer correlation function}
Next let us consider the dimer-dimer correlation function defined as 
\begin{align}
D_n\equiv\dimer{n}{n+1}=3\zcorf{1}{2}{n}{n+1}{z}{z}{z}{z}+6\zcorf{1}{2}{n}{n+1}{x}{x}{z}{z},
\end{align}
which should asymptotically coincide with the square of the ground state energy 
\begin{align}
\lim_{n\to\infty}D_n=e_0^2=\(\frac14-\za{1}\)^2=0.1963794236382287\cdots.
\end{align}
From our results of all the density matrix elements for $n=6$, 
we can exactly evaluate the dimer-dimer correlation functions $D_n$ for up to $n=5$: 
\begin{align}
D_3-e_0^2&=\frac56\za{3}-\frac43\za{1}\za{3}-\za{3}^2-\frac56\za{5}+\frac{10}3\za{1}\za{5}
-\za{1}^2\nn\\&=0.06082478294036410\cdots,\db
D_4-e_0^2&=\frac56\za{3}-\frac{14}3\za{1}\za{3}-\frac{43}3\za{3}^2-\frac{15}4\za{5}+\frac{160}3\za{1}\za{5}-5\za{3}\za{5}\nn\\&-\frac{50}3\za{5}^2+\frac{35}{12}\za{7}-\frac{140}3\za{1}\za{7}+35\za{3}\za{7}-\za{1}^2\nn\\&=-0.02773785800119889\cdots,\db
D_5-e_0^2&=\frac56\za{3}-10\za{1}\za{3}-\frac{221}3\za{3}^2-\frac{320}9\za{3}^2-\frac{263}{30}\za{5}+\frac{862}3\za{1}\za{5}\nn\db&-\frac{196}5\za{3}\za{5}+\frac{640}3\za{1}\za{3}\za{5}-\frac{904}{15}\za{3}^2\za{5}-1381\za{5}^2\nn\db&+\frac{3616}3\za{1}\za{5}^2+\frac{680}3\za{3}\za{5}^2+\frac{1040}9\za{5}^3+\frac{161}6\za{7}-\frac{3220}3\za{1}\za{7}\nn\db&+3038\za{3}\za{7}-\frac{6328}3\za{1}\za{3}\za{7}-\frac{1904}3\za{3}^2\za{7}+\frac{1204}3\za{5}\za{7}\nn\db&-\frac{1456}3\za{3}\za{5}\za{7}-\frac{3773}3\za{7}^2+\frac{5096}3\za{1}\za{7}^2-\frac{189}{10}\za{9}+798\za{1}\za{9}\nn\db&-\frac{14224}5\za{3}\za{9}+1904\za{1}\za{3}\za{9}+\frac{4368}5\za{3}^2\za{9}+2156\za{5}\za{9}\nn\db&-2912\za{1}\za{5}\za{9}-\za{1}^2
\nn\\&=0.01892813120084483\cdots,
\end{align}
The difference $D_n-e_0^2$ also should decay algebraicly. 
\subsection{$AP(n)$: a probability to find an antiferromagnetic string of length $n$}
Emptiness Formation Probability $P(n)$ (\ref{efp}) represents 
the probability to find a ferromagnetic string of length $n$. 
Similarly we may consider a probability to find an antiferromagnetic string of length $n$, 
which we denote by $AP(n)$
\begin{align}
AP(n)\equiv P^{+-+-\cdots}_{+-+-\cdots}+P^{-+-+\cdots}_{-+-+\cdots}.
\end{align}
From our results we can exactly evaluate the $AP(n)$ for up to $n=6$
\begin{align}
AP(2)&=P^{+-}_{+-}+P^{-+}_{-+}=\frac{1}{3}+\frac{2}{3}\za{1}=0.7954314537066302\cdots,\db
AP(3)&=P^{+-+}_{+-+}+P^{-+-}_{-+-}=\frac{1}{6}-\frac{2}{3}\za{1}+\za{3}=0.6061112236630655\cdots,\db
AP(4)&=P^{+-+-}_{+-+-}+P^{-+-+}_{-+-+}\nn\\&=\frac{1}{15}-\frac{8}{15}\za{3}+\frac{4}{3}\za{1}\za{3}+\frac{2}{5}\za{3}^2+\frac{2}{3}\za{5}-\frac{4}{3}\za{1}\za{5}
\nn\\&=0.4938083479102196\cdots,\db
AP(5)&=P^{+-+-+}_{+-+-+}+P^{-+-+-}_{-+-+-}\nn\\&=\frac{1}{30}-\frac{2}{3}\za{1}+\frac{91}{15}\za{3}-\frac{124}{9}\za{1}\za{3}-\frac{134}{5}\za{3}^2-\frac{355}{18}\za{5}+\frac{328}{3}\za{1}\za{5}\nn\\&-\frac{38}{9}\za{3}\za{5}-\frac{250}{9}\za{5}^2+\frac{259}{18}\za{7}-\frac{854}{9}\za{1}\za{7}+\frac{175}{3}\za{3}\za{7}\nn\\&
=0.3943245947356898\cdots,\db
AP(6)&=P^{+-+-+-}_{+-+-+-}+P^{-+-+-+}_{-+-+-+}\nn\\&=\frac{1}{70}-\frac{1}{5}\za{1}+\frac{44}{45}\za{3}+\frac{36}{5}\za{1}\za{3}+\frac{2566}{45}\za{3}^2\nn\db&+\frac{4072}{135}\za{3}^3+\frac{9754}{1575}\za{5}-\frac{10984}{45}\za{1}\za{5}-\frac{122}{75}\za{3}\za{5}\nn\db&-\frac{8144}{45}\za{1}\za{3}\za{5}+\frac{1368}{25}\za{3}^2\za{5}+\frac{374128}{315}\za{5}^2-\frac{5472}{5}\za{1}\za{5}^2\nn\db&-\frac{1868}{9}\za{3}\za{5}^2-\frac{19868}{189}\za{5}^3-\frac{1936}{45}\za{7}+\frac{50848}{45}\za{1}\za{7}\nn\db&-\frac{39794}{15}\za{3}\za{7}+\frac{9576}{5}\za{1}\za{3}\za{7}+\frac{26152}{45}\za{3}^2\za{7}-\frac{1868}{5}\za{5}\za{7}\nn\db&+\frac{19868}{45}\za{3}\za{5}\za{7}+\frac{47299}{45}\za{7}^2-\frac{69538}{45}\za{1}\za{7}^2+\frac{902}{25}\za{9}\nn\db&-\frac{4464}{5}\za{1}\za{9}+\frac{190052}{75}\za{3}\za{9}-\frac{26152}{15}\za{1}\za{3}\za{9}\nn\db&-\frac{19868}{25}\za{3}^2\za{9}-\frac{27028}{15}\za{5}\za{9}+\frac{39736}{15}\za{1}\za{5}\za{9}
\nn\\&=0.3239037769698205\cdots.
\end{align}

It is known that the EFP $P(n)$ shows a Gaussian decay \cite{KLNS, KMST}
\begin{align}
P(n)\simeq An^{-\gamma}C^{-n^2}. 
\end{align}

On the other hand, the analytical asymptotic behavior of $AP(n)$ has not been studied. 
However, Y. Nishiyama noticed that it decays roughly $AP(n)\simeq\[AP(2)\]^{n-1}$ 
from his numerical data before \cite{Nishiyama}. 
We compare the formula with our exact data in Table \ref{apn}. 
\begin{table}[h]
\caption{Asymptotic behavior of $AP(n)$}
\label{apn}
\begin{center}
\begin{tabular}
{@{\hspace{\tabcolsep}\extracolsep{\fill}}ccc} 
\hline
 $n$ &    $AP(n)$ Exact     & $\[AP(2)\]^{n-1}$    \\
\hline
  2  & 0.795431 & 0.795431 \\
  3  & 0.606111 & 0.632711 \\
  4  & 0.493808 & 0.503278 \\
  5  & 0.394325 & 0.400323 \\
  6  & 0.323904 & 0.318430 \\
\hline
\end{tabular}
\end{center}
\end{table}
\section{Reduced density matrix and entanglement entropy}
\begin{figure}
\begin{center}
\includegraphics[width=0.7\textwidth]{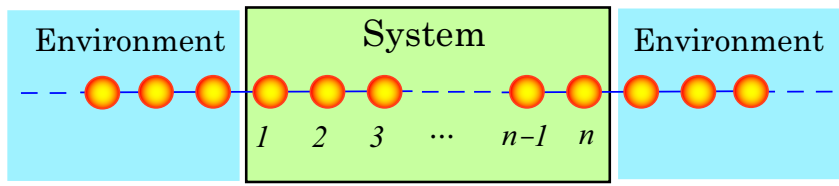}
\caption{Finite sub-chain of length $n$ in the infinite chain}
\label{rd}
\end{center}
\end{figure}

\begin{figure}
\includegraphics[width=0.9\textwidth]{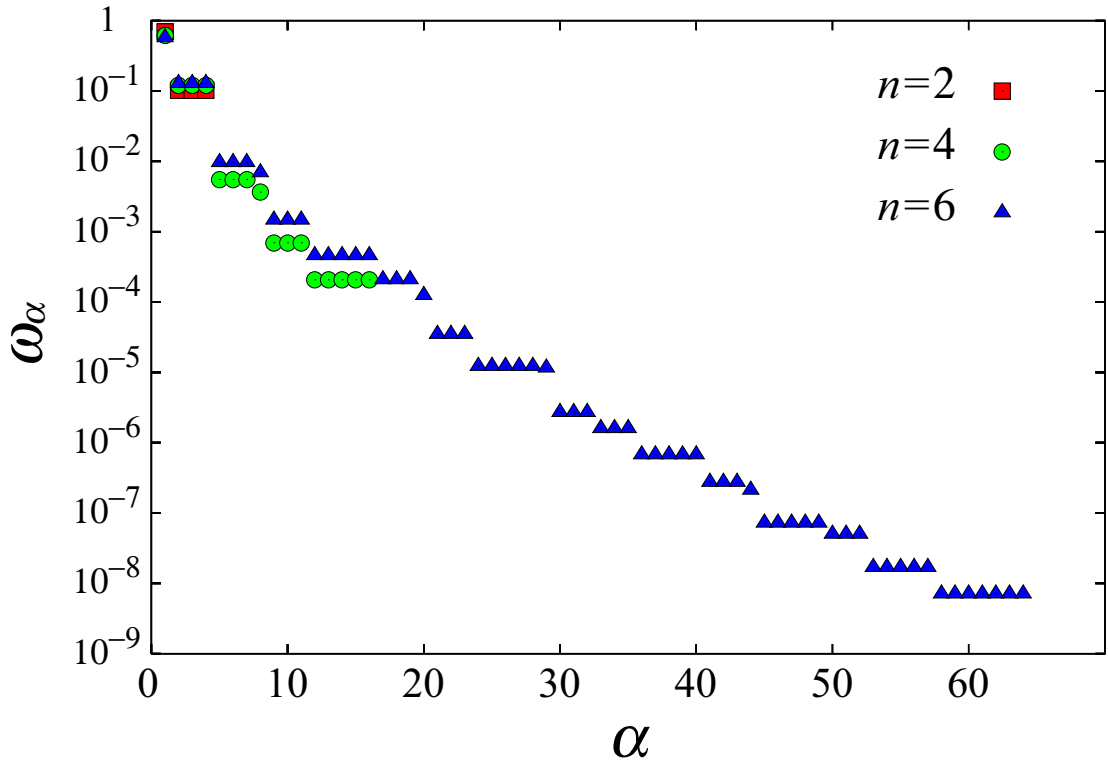}
\includegraphics[width=0.9\textwidth]{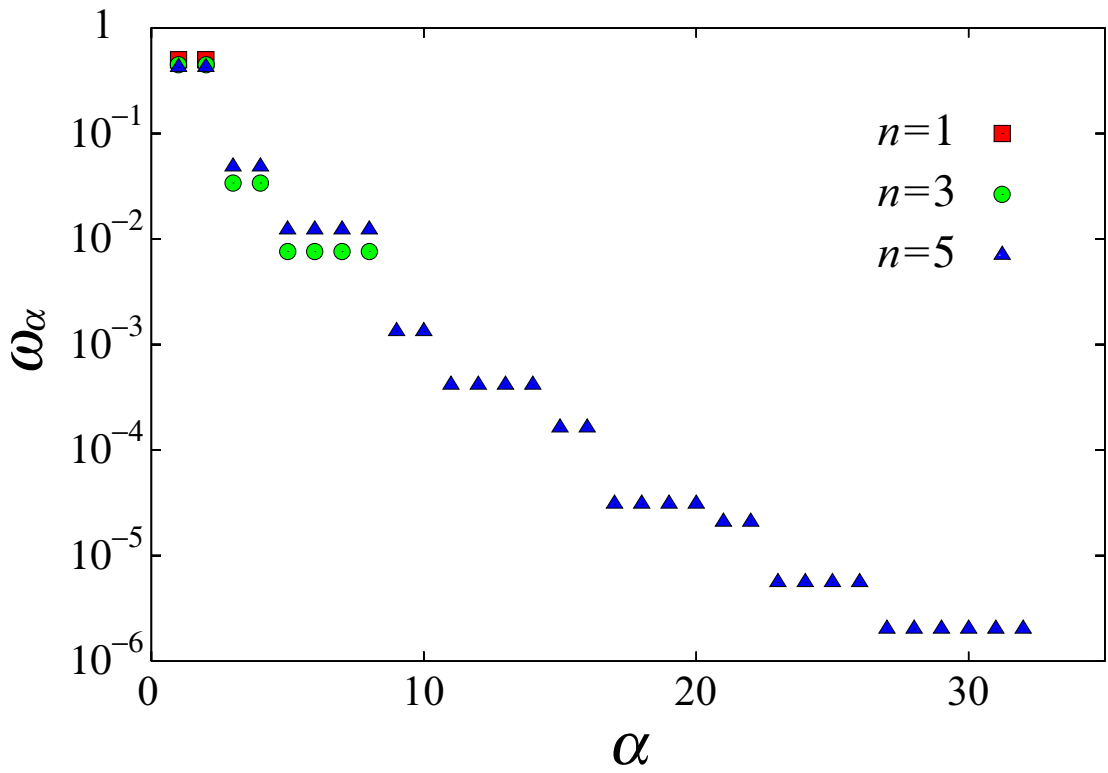}
\caption{Eigenvalue-distribution of density matrices}
\label{ev}
\end{figure}

\setcounter{equation}{0}
At zero temperature, the density matrix for the infinite system 
can be written as 
\begin{align}
\rho_{\rm T}\equiv|{\rm GS}\ket\bra{\rm GS}|,
\end{align}
where $|{\rm GS}\ket$ denotes the antiferromagnetic ground state. 
Let us consider a finite sub-chain of length $n$ in the infinite chain, 
the rest of which can be regarded as the environment (Figure \ref{rd}). 
The reduced density matrix for this sub-chain is obtained by tracing out the environment 
from the infinite chain:
\begin{align}
\rho_n\equiv{\rm tr}_E\rho_{\rm T}=\[P^{\epDs{1}{n}}_{\eps{1}{n}}\]_{\ep{j},\epD{j}=\pm}. 
\end{align}
From our results, we have computed all the eigenvalues 
$\omega_{\alpha}$ $(\alpha=1,2,\cdots,2^n)$ of the reduced density matrix $\rho_n$ 
for up to $n=6$, which are shown in Figure \ref{ev}. 

We have found that the smallest eigenvalue $\omega_{2^n}$ is, for any $n$, 
the emptiness formation probability $P(n)$. 
This is $(n+1)$-fold degenerate since the reduced density matrix has the block-diagonalized 
form by the magnetization of the sub-chain $M_{z}=\sum_{j=1}^n\epD{j}=\sum_{j=1}^n\ep{j}$, 
each of which has the non-degenerate smallest eigenvalue $P(n)$. 
Or in other words, it is a consequence of $SU(2)$ invariance of the Hamiltonian. 

\begin{figure}
\includegraphics[width=0.9\textwidth]{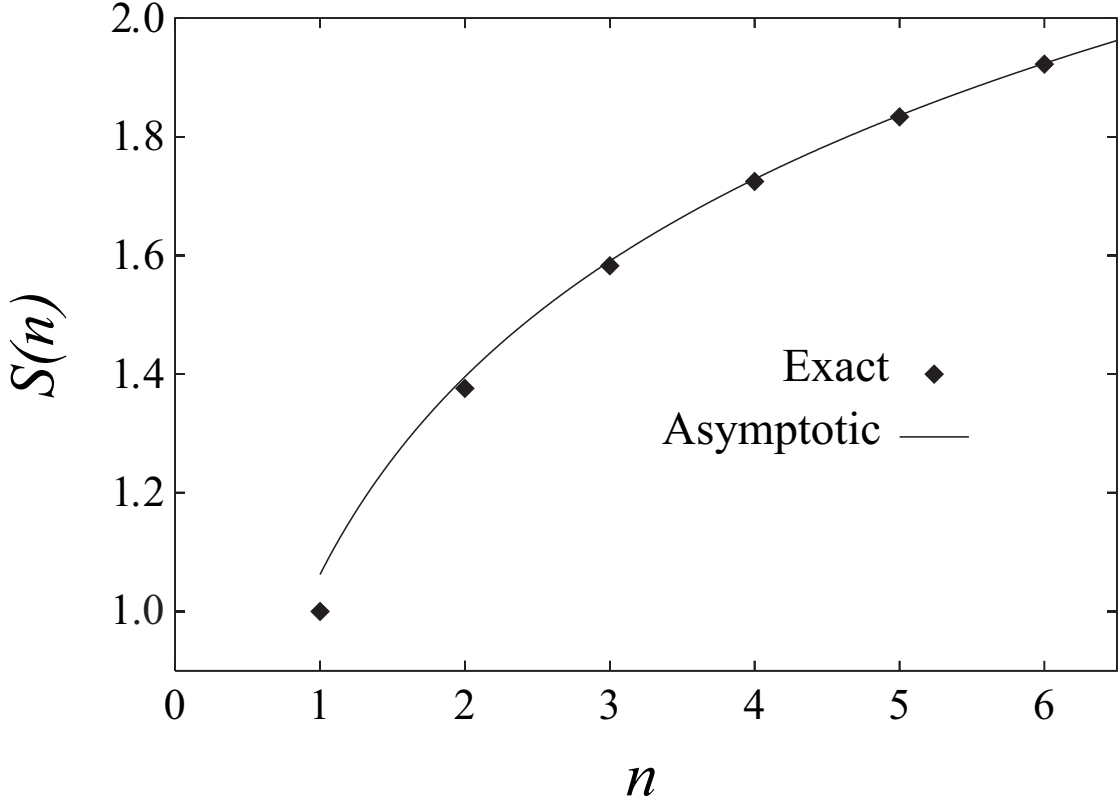}
\caption{von Neumann entropy $S(n)$ of a finite sub-chain of length $n$}
\label{ent}
\end{figure}

From these results we can calculate the von Neumann entropy (Entanglement entropy). 
\begin{align}
S(n)\equiv-{\rm tr}\rho_n\log_2\rho_n
=-\sum_{\alpha=1}^{2^n}\omega_{\alpha}\log_2\omega_{\alpha},
\end{align}
which is considered to measure how the ground state is entangled \cite{VLRK}. 
The exact numerical values of $S(n)$ up to $n=6$ are shown in Table \ref{von}. 
\begin{table}
\begin{center}
\caption{von Neumann entropy $S(n)$ of a finite sub-chain of length $n$}
\label{von}
\begin{tabular}
{@{\hspace{\tabcolsep}\extracolsep{\fill}}cccc} 
\hline
$S$(1)&$S$(2)&$S$(3)&$S$(4)\\
\hline
1&1.3758573262887466&1.5824933209573855&1.7247050949099274\\
\hline
\end{tabular}
\end{center}
\begin{center}
\begin{tabular}
{@{\hspace{\tabcolsep}\extracolsep{\fill}}cc} 
\hline
$S$(5)&$S$(6)\\
\hline
1.833704916848315&1.922358833819333\\
\hline
\end{tabular}
\end{center}
\end{table}
It can be seen that as $n$ grows the von Neumann entropy is increasing in small steps. 
The asymptotic behavior of this entropy is discussed in \cite{VLRK}. 
In the massive region $\Delta>1$, the von Neumann entropy will be saturated as $n$ grows, 
which means the ground state is well approximated by a subsystem of a finite length 
corresponding to the large eigenvalues of reduced density matrix. 
On the other hand, in the massless case $-1<\Delta\leq1$, 
the conformal field theory predict that 
the von Neumann entropy shows a logarithmic divergence \cite{HLW}, 
of which the explicit form for the XXX case $\Delta=1$ reads 
\begin{align}
S(n)\simeq\frac13\log_2n+k.
\end{align}
We estimate the numerical value of the constant term $k=1.06209$
by the extrapolation $S(n)-\frac13\log_2n=c_0+c_1/n+c_2/n^2$. 
Our exact results for up to $n=6$ agree quite well with the asymptotic formula 
as shown in Figure \ref{ent}. 

%
\section{Summary and discussion}
We have succeeded in obtaining the analytical forms of all the density matrix elements 
on six lattice sites for Heisenberg chain using the algebraic method based on the qKZ 
equations developed in \cite{BST}. 
We have obtained several physically interesting correlation functions, 
such as chiral correlation functions, dimer-dimer correlation functions, etc$\cdots$.
Subsequently we have calculated all the eigenvalues of the reduced density matrix 
$\rho_n$ for up to $n=6$. 
We observe that the smallest eigenvalue is the emptiness formation probability $P(n)$ 
and is $(n+1)$-fold degenerate. 
Of course, it may be more desirable if we could describe the eigenvector corresponding to 
the largest eigenvalue. 
From these results we have computed the von Neumann entropy, 
which shows a good agreement with the asymptotic formula derived via the conformal field 
theory. 
%
\section*{Acknowledgement}
The authors are grateful to H. Boos, V.E. Korepin, Y. Nishiyama and K. Sakai for 
fruitful collaborations of early stage of this work. 
Especially we thank V.E. Korepin for suggesting us to calculate the entanglement entropy. 
We also thank M. Bortz and M. Batchelor for valuable discussions. 
This work is in part supported by Grant-in-Aid for the Scientific Research (B) No. 18340112. 
from the Ministry of Education, Culture, Sports, Science and Technology, Japan. 
It is also supported by JSPS and ARC under the Japan-Australian joint project in 
cooperative science program. 
\begin{appendix}
%
\section{Known results for up to $n=5$}
In this appendix we present all the independent correlation functions for up to $n=5$ 
except for the two-point correlation functions $\zcor{1}{n}$ given in Introduction. 
First we give the correlation functions on four lattice sites \cite{SSNT}: 
\begin{align}
\zcorf{1}{2}{3}{4}{x}{x}{z}{z}=&
\frac{1}{240}+\frac{1}{12}\za{1}-\frac{91}{180}\za{3}+\frac{2}{9}\za{1}\za{3}+\frac{1}{15}\za{3}^2+\frac{7}{18}\za{5}-\frac{2}{9}\za{1}\za{5}\nn\\
=&0.02750969925180030\cdots,\db
\zcorf{1}{2}{3}{4}{x}{z}{x}{z}=&
\frac{1}{240}-\frac{1}{6}\za{1}+\frac{77}{90}\za{3}-\frac{5}{9}\za{1}\za{3}-\frac{4}{15}\za{3}^2-\frac{13}{18}\za{5}+\frac{8}{9}\za{1}\za{5}\nn\\
=&-0.007080326815007911\cdots,\db
\zcorf{1}{2}{3}{4}{x}{z}{z}{x}=&
\frac{1}{240}-\frac{1}{4}\za{1}+\frac{169}{180}\za{3}-\frac{5}{9}\za{1}\za{3}-\frac{4}{15}\za{3}^2-\frac{13}{18}\za{5}+\frac{8}{9}\za{1}\za{5}\nn\\
=&0.01028596458580462\cdots,\db
\zcorf{1}{2}{3}{4}{z}{z}{z}{z}=&\zcorf{1}{2}{3}{4}{x}{x}{z}{z}
+\zcorf{1}{2}{3}{4}{x}{z}{x}{z}+\zcorf{1}{2}{3}{4}{x}{z}{z}{x}. 
\end{align}
Note that on the antiferromagnetic ground state without magnetic field, 
the correlation functions with an odd number of spin operators vanish. 
Considering the isotropy of the Hamiltonian (\ref{Ham}), 
we can see that the independent correlation functions on four lattice sites are 
restricted to the above ones and the two-point correlation function $\zcor{1}{4}$. 

Correspondingly the independent correlation functions on five lattice sites are 
\cite{BST} 
\begin{align}
\zcorf{1}{2}{3}{5}{x}{x}{z}{z}=&
\frac{1}{240}+\frac{1}{12}\za{1}-\frac{517}{360}\za{3}+\frac{25}{9}\za{1}\za{3}+\frac{203}{45}\za{3}^2+\frac{305}{72}\za{5} \nn\\& -\frac{172}{9}\za{1}\za{5}+\frac{4}{9}\za{3}\za{5}+\frac{40}{9}\za{5}^2-\frac{35}{12}\za{7}+\frac{49}{3}\za{1}\za{7} \nn\\& -\frac{28}{3}\za{3}\za{7} 
= -0.009892435084700294 \cdots, \db
\zcorf{1}{2}{3}{5}{x}{z}{x}{z}=&
\frac{1}{240}-\frac{1}{4}\za{1}+\frac{301}{120}\za{3}-6\za{1}\za{3}-\frac{1079}{90}\za{3}^2-\frac{65}{8}\za{5} \nn\\& +\frac{146}{3}\za{1}\za{5}-\frac{37}{18}\za{3}\za{5}-\frac{110}{9}\za{5}^2+\frac{35}{6}\za{7}-42\za{1}\za{7} \nn\\& +\frac{77}{3}\za{3}\za{7} 
= 0.002788973399542967 \cdots, \db
\zcorf{1}{2}{3}{5}{x}{z}{z}{x}=&
\frac{1}{240}-\frac{5}{12}\za{1}+\frac{569}{180}\za{3}-\frac{61}{9}\za{1}\za{3}-\frac{1109}{90}\za{3}^2-\frac{155}{18}\za{5} \nn\\& +\frac{448}{9}\za{1}\za{5}-\frac{37}{18}\za{3}\za{5}-\frac{110}{9}\za{5}^2+\frac{35}{6}\za{7}-42\za{1}\za{7} \nn\\& +\frac{77}{3}\za{3}\za{7} 
= -0.005056660894819286 \cdots, \db
\zcorf{1}{2}{3}{5}{z}{z}{z}{z}=&
\zcorf{1}{2}{3}{5}{x}{x}{z}{z}+\zcorf{1}{2}{3}{5}{x}{z}{x}{z}
+\zcorf{1}{2}{3}{5}{x}{z}{z}{x}, \db
\zcorf{1}{2}{4}{5}{x}{x}{z}{z}=&
\frac{1}{240}+\frac{1}{4}\za{1}-\frac{419}{120}\za{3}+\frac{70}{9}\za{1}\za{3}
+\frac{221}{15}\za{3}^2+\frac{889}{72}\za{5} \nn\\& -\frac{184}{3}\za{1}\za{5}
+\frac{17}{9}\za{3}\za{5}+\frac{130}{9}\za{5}^2-\frac{329}{36}\za{7}
+\frac{476}{9}\za{1}\za{7} \nn\\& -\frac{91}{3}\za{3}\za{7}
 = 0.01857662093837095 \cdots, \db
\zcorf{1}{2}{4}{5}{x}{z}{x}{z}=&
\frac{1}{240}-\frac{5}{12}\za{1}+\frac{1883}{360}\za{3}-\frac{112}{9}\za{1}\za{3}
-\frac{1102}{45}\za{3}^2-\frac{457}{24}\za{5} \nn\\& +\frac{908}{9}\za{1}\za{5}
-\frac{11}{3}\za{3}\za{5}-\frac{220}{9}\za{5}^2+\frac{511}{36}\za{7}
-\frac{784}{9}\za{1}\za{7} \nn\\& +\frac{154}{3}\za{3}\za{7}
 = -0.001089368972914637 \cdots, \db
\zcorf{1}{2}{4}{5}{x}{z}{z}{x}=&
\frac{1}{240}-\frac{1}{2}\za{1}+\frac{497}{90}\za{3}-\frac{112}{9}\za{1}\za{3}
-\frac{1102}{45}\za{3}^2-\frac{77}{4}\za{5} \nn\\& +\frac{908}{9}\za{1}\za{5}
-\frac{11}{3}\za{3}\za{5}-\frac{220}{9}\za{5}^2+\frac{511}{36}\za{7}
-\frac{784}{9}\za{1}\za{7} \nn\\& +\frac{154}{3}\za{3}\za{7}
 = 0.001573361370145065 \cdots, \db
\zcorf{1}{2}{4}{5}{z}{z}{z}{z}=&
\zcorf{1}{2}{4}{5}{x}{x}{z}{z}+\zcorf{1}{2}{4}{5}{x}{z}{x}{z}
+\zcorf{1}{2}{4}{5}{x}{z}{z}{x}. 
\end{align}
\section{Explicit form for all the correlation functions on $6$ lattice sites}
\setcounter{equation}{0}
In this appendix the explicit results for 
all the correlation functions on $6$ lattice sites are shown. 
There are $24$ independent correlation functions among them. 
Any other correlation functions can be computed by the following relations: 

\begin{align}
&\zcorf{1}{3}{4}{6}{z}{z}{z}{z}=\zcorf{1}{3}{4}{6}{x}{x}{z}{z}+\zcorf{1}{3}{4}{6}{x}{z}{x}{z}
+\zcorf{1}{3}{4}{6}{x}{z}{z}{x}, \\[5pt]&
\zcorf{1}{2}{5}{6}{z}{z}{z}{z}=\zcorf{1}{2}{5}{6}{x}{x}{z}{z}+\zcorf{1}{2}{5}{6}{x}{z}{x}{z}
+\zcorf{1}{2}{5}{6}{x}{z}{z}{x},\\[5pt]&
\zcorf{1}{4}{5}{6}{z}{z}{z}{z}=\zcorf{1}{4}{5}{6}{x}{x}{z}{z}+\zcorf{1}{4}{5}{6}{x}{z}{x}{z}
+\zcorf{1}{4}{5}{6}{x}{z}{z}{x},\\[5pt]&
\zcorf{1}{3}{5}{6}{z}{z}{z}{z}=\zcorf{1}{3}{5}{6}{x}{x}{z}{z}+\zcorf{1}{3}{5}{6}{x}{z}{x}{z}
+\zcorf{1}{3}{5}{6}{x}{z}{z}{x},
\end{align}
\begin{align}
&\zcors{x}{x}{y}{y}{z}{z}=\zcors{x}{x}{x}{x}{y}{y}
-\zcors{x}{x}{y}{z}{y}{z}-\zcors{x}{x}{y}{z}{z}{y},\\[5pt]&
\zcors{x}{y}{x}{z}{y}{z}=\zcors{x}{x}{x}{y}{x}{y}
-\zcors{x}{x}{y}{z}{y}{z}-\zcors{x}{y}{x}{z}{z}{y},\\[5pt]&
\zcors{x}{y}{y}{z}{z}{x}=\zcors{x}{x}{x}{y}{y}{x}
-\zcors{x}{x}{y}{z}{z}{y}-\zcors{x}{y}{x}{z}{z}{y},\\[5pt]&
\zcors{x}{y}{z}{x}{y}{z}=\zcors{x}{x}{y}{x}{x}{y}
-\zcors{x}{x}{y}{z}{z}{y}-\zcors{x}{y}{z}{x}{z}{y},\\[5pt]&
\zcors{x}{y}{z}{y}{z}{x}=\zcors{x}{x}{y}{x}{y}{x}
-\zcors{x}{x}{y}{z}{y}{z}-\zcors{x}{y}{z}{x}{z}{y},\\[5pt]&
\zcors{x}{y}{z}{z}{x}{y}=\zcors{x}{y}{x}{x}{x}{y}
-\zcors{x}{y}{x}{z}{z}{y}-\zcors{x}{y}{z}{x}{z}{y},
\end{align}
\begin{align}
&\zcors{x}{x}{y}{y}{x}{x}=\zcors{x}{y}{z}{z}{y}{x}
+\zcors{x}{x}{x}{x}{y}{y}+\zcors{x}{y}{x}{x}{x}{y}\nn\\&
-\zcors{x}{x}{y}{z}{y}{z}-\zcors{x}{x}{y}{z}{z}{y}
-\zcors{x}{y}{x}{z}{z}{y}-\zcors{x}{y}{z}{x}{z}{y},\\[5pt]&
\zcors{x}{y}{x}{x}{y}{x}=\zcors{x}{y}{z}{z}{y}{x}
+\zcors{x}{x}{x}{y}{x}{y}+\zcors{x}{x}{y}{x}{x}{y}\nn\\&
-\zcors{x}{x}{y}{z}{y}{z}-\zcors{x}{x}{y}{z}{z}{y}
-\zcors{x}{y}{x}{z}{z}{y}-\zcors{x}{y}{z}{x}{z}{y},\\[5pt]&
\zcors{x}{y}{y}{y}{y}{x}=\zcors{x}{y}{z}{z}{y}{x}
+\zcors{x}{x}{x}{y}{y}{x}+\zcors{x}{x}{y}{x}{y}{x}\nn\\&
-\zcors{x}{x}{y}{z}{y}{z}-\zcors{x}{x}{y}{z}{z}{y}
-\zcors{x}{y}{x}{z}{z}{y}-\zcors{x}{y}{z}{x}{z}{y},
\end{align}
\begin{align}
&\zcors{x}{x}{x}{x}{x}{x}=\zcors{x}{y}{z}{z}{y}{x}
+\zcors{x}{x}{x}{x}{y}{y}+\zcors{x}{x}{x}{y}{x}{y}\nn\\&
+\zcors{x}{x}{x}{y}{y}{x}+\zcors{x}{x}{y}{x}{x}{y}
+\zcors{x}{x}{y}{x}{y}{x}+\zcors{x}{y}{x}{x}{x}{y}\nn\\&
-\zcors{x}{x}{y}{z}{y}{z}-\zcors{x}{x}{y}{z}{z}{y}
-\zcors{x}{y}{x}{z}{z}{y}-\zcors{x}{y}{z}{x}{z}{y}.
\end{align}

Below we shall give the explicit results for the independent ones. 

\begin{align}
\zcor{1}{6}&=
\frac{1}{12}-\frac{25}{3}\za{1}+\frac{800}{9}\za{3}-\frac{1192}{3}\za{1}\za{3}-\frac{15368}{9}\za{3}^2\nn\\&-608\za{3}^3-\frac{4228}{9}\za{5}+\frac{64256}{9}\za{1}\za{5}-\frac{976}{9}\za{3}\za{5}\nn\\&+3648\za{1}\za{3}\za{5}-\frac{3328}{3}\za{3}^2\za{5}-\frac{76640}{3}\za{5}^2
\nn\\&+\frac{66560}{3}\za{1}\za{5}^2+\frac{12640}{3}\za{3}\za{5}^2+\frac{6400}{3}\za{5}^3+\frac{9674}{9}\za{7}\nn\\&+56952\za{3}\za{7}-\frac{225848}{9}\za{1}\za{7}-\frac{116480}{3}\za{1}\za{3}\za{7}\nn\\&-\frac{35392}{3}\za{3}^2\za{7}+7840\za{5}\za{7}-8960\za{3}\za{5}\za{7}\nn\\&-\frac{66640}{3}\za{7}^2+31360\za{1}\za{7}^2-686\za{9}\nn\\&+18368\za{1}\za{9}-53312\za{3}\za{9}+35392\za{1}\za{3}\za{9}\nn\\&+16128\za{3}^2\za{9}+38080\za{5}\za{9}-53760\za{1}\za{5}\za{9}\nn\\&
=-0.03089036664760932\cdots
\end{align}

\begin{align}
\zcorf{1}{3}{4}{6}{x}{x}{z}{z}&=
\frac{1}{240}+\frac{1}{10}\za{1}-\frac{259}{90}\za{3}+\frac{463}{45}\za{1}\za{3}+\frac{1951}{45}\za{3}^2\nn\\&+\frac{2404}{135}\za{3}^3+\frac{1348}{75}\za{5}-\frac{8918}{45}\za{1}\za{5}-\frac{3127}{225}\za{3}\za{5}\nn\\&-\frac{4808}{45}\za{1}\za{3}\za{5}+\frac{7804}{225}\za{3}^2\za{5}+\frac{33598}{45}\za{5}^2\nn\\&-\frac{31216}{45}\za{1}\za{5}^2-\frac{1196}{9}\za{3}\za{5}^2-\frac{1808}{27}\za{5}^3-\frac{413}{9}\za{7}\nn\\&+\frac{36421}{45}\za{1}\za{7}-\frac{8393}{5}\za{3}\za{7}+\frac{54628}{45}\za{1}\za{3}\za{7}\nn\\&+\frac{16744}{45}\za{3}^2\za{7}-\frac{10864}{45}\za{5}\za{7}+\frac{12656}{45}\za{3}\za{5}\za{7}\nn\\&+\frac{29008}{45}\za{7}^2-\frac{44296}{45}\za{1}\za{7}^2+\frac{1533}{50}\za{9}\nn\\&-\frac{3108}{5}\za{1}\za{9}+\frac{120344}{75}\za{3}\za{9}-\frac{16744}{15}\za{1}\za{3}\za{9}\nn\\&-\frac{12656}{25}\za{3}^2\za{9}-\frac{16576}{15}\za{5}\za{9}+\frac{25312}{15}\za{1}\za{5}\za{9}\nn\\&
=0.003681507672875026\cdots
\end{align}

\begin{align}
\zcorf{1}{3}{4}{6}{x}{z}{x}{z}&=
\frac{1}{240}-\frac{19}{60}\za{1}+\frac{1007}{180}\za{3}-\frac{1252}{45}\za{1}\za{3}-\frac{1978}{15}\za{3}^2\nn\\&-\frac{6256}{135}\za{3}^3-\frac{979}{25}\za{5}+\frac{25322}{45}\za{1}\za{5}+\frac{538}{225}\za{3}\za{5}\nn\\&+\frac{12512}{45}\za{1}\za{3}\za{5}-\frac{6292}{75}\za{3}^2\za{5}-\frac{85687}{45}\za{5}^2\nn\\&+\frac{25168}{15}\za{1}\za{5}^2+\frac{2864}{9}\za{3}\za{5}^2+\frac{4352}{27}\za{5}^3+\frac{1771}{18}\za{7}\nn\\&-\frac{89929}{45}\za{1}\za{7}+\frac{191513}{45}\za{3}\za{7}-\frac{44044}{15}\za{1}\za{3}\za{7}\nn\\&-\frac{40096}{45}\za{3}^2\za{7}+\frac{26866}{45}\za{5}\za{7}-\frac{30464}{45}\za{3}\za{5}\za{7}\nn\\&-\frac{73402}{45}\za{7}^2+\frac{106624}{45}\za{1}\za{7}^2-\frac{3227}{50}\za{9}\nn\\&+\frac{7322}{5}\za{1}\za{9}-\frac{298886}{75}\za{3}\za{9}+\frac{40096}{15}\za{1}\za{3}\za{9}\nn\\&+\frac{30464}{25}\za{3}^2\za{9}+\frac{41944}{15}\za{5}\za{9}-\frac{60928}{15}\za{1}\za{5}\za{9}\nn\\&
=-0.001116347734065082\cdots
\end{align}

\begin{align}
\zcorf{1}{3}{4}{6}{x}{z}{z}{x}&=
\frac{1}{240}-\frac{13}{20}\za{1}+\frac{479}{60}\za{3}-\frac{1537}{45}\za{1}\za{3}-\frac{6709}{45}\za{3}^2\nn\\&-\frac{2452}{45}\za{3}^3-\frac{10201}{225}\za{5}+\frac{9524}{15}\za{1}\za{5}+\frac{37}{25}\za{3}\za{5}\nn\\&+\frac{4904}{15}\za{1}\za{3}\za{5}-\frac{22616}{225}\za{3}^2\za{5}-\frac{102592}{45}\za{5}^2\nn\\&+\frac{90464}{45}\za{1}\za{5}^2+\frac{1148}{3}\za{3}\za{5}^2+\frac{1744}{9}\za{5}^3+\frac{973}{9}\za{7}\nn\\&-\frac{103544}{45}\za{1}\za{7}+\frac{229138}{45}\za{3}\za{7}-\frac{158312}{45}\za{1}\za{3}\za{7}\nn\\&-\frac{16072}{15}\za{3}^2\za{7}+\frac{31906}{45}\za{5}\za{7}-\frac{12208}{15}\za{3}\za{5}\za{7}\nn\\&-\frac{9898}{5}\za{7}^2+\frac{42728}{15}\za{1}\za{7}^2-\frac{3507}{50}\za{9}\nn\\&+\frac{8512}{5}\za{1}\za{9}-\frac{119742}{25}\za{3}\za{9}+\frac{16072}{5}\za{1}\za{3}\za{9}\nn\\&+\frac{36624}{25}\za{3}^2\za{9}+\frac{16968}{5}\za{5}\za{9}-\frac{24416}{5}\za{1}\za{5}\za{9}\nn\\&
=0.003069653070471227\cdots
\end{align}

\begin{align}
\zcorf{1}{2}{5}{6}{x}{x}{z}{z}&=
\frac{1}{240}+\frac{29}{60}\za{1}-\frac{568}{45}\za{3}+\frac{1024}{15}\za{1}\za{3}+\frac{15863}{45}\za{3}^2\nn\\&+\frac{5872}{45}\za{3}^3+\frac{25486}{225}\za{5}-\frac{69922}{45}\za{1}\za{5}-\frac{13598}{225}\za{3}\za{5}\nn\\&-\frac{11744}{15}\za{1}\za{3}\za{5}+\frac{53816}{225}\za{3}^2\za{5}+\frac{246547}{45}\za{5}^2\nn\\&-\frac{215264}{45}\za{1}\za{5}^2-\frac{2728}{3}\za{3}\za{5}^2-\frac{4144}{9}\za{5}^3-\frac{61327}{180}\za{7}\nn\\&+\frac{271264}{45}\za{1}\za{7}-\frac{61474}{5}\za{3}\za{7}+\frac{376712}{45}\za{1}\za{3}\za{7}\nn\\&+\frac{38192}{15}\za{3}^2\za{7}-\frac{79436}{45}\za{5}\za{7}+\frac{29008}{15}\za{3}\za{5}\za{7}\nn\\&+\frac{23373}{5}\za{7}^2-\frac{101528}{15}\za{1}\za{7}^2+\frac{11977}{50}\za{9}\nn\\&-\frac{22722}{5}\za{1}\za{9}+\frac{290752}{25}\za{3}\za{9}-\frac{38192}{5}\za{1}\za{3}\za{9}\nn\\&-\frac{87024}{25}\za{3}^2\za{9}-\frac{40068}{5}\za{5}\za{9}+\frac{58016}{5}\za{1}\za{5}\za{9}\nn\\&
=0.02384723373803033\cdots
\end{align}

\begin{align}
\zcorf{1}{2}{5}{6}{x}{z}{x}{z}&=
\frac{1}{240}-\frac{23}{30}\za{1}+\frac{563}{30}za{3}-\frac{4678}{45}\za{1}\za{3}-\frac{8114}{15}\za{3}^2\nn\\&-\frac{27224}{135}\za{3}^3-\frac{38389}{225}\za{5}+\frac{35666}{15}\za{1}\za{5}+\frac{18902}{225}\za{3}\za{5}\nn\\&+\frac{54448}{45}\za{1}\za{3}\za{5}-\frac{82984}{225}\za{3}^2\za{5}-\frac{42242}{5}\za{5}^2\nn\\&+\frac{331936}{45}\za{1}\za{5}^2+\frac{12616}{9}\za{3}\za{5}^2+\frac{19168}{27}\za{5}^3+\frac{23177}{45}\za{7}\nn\\&-\frac{414911}{45}\za{1}\za{7}+\frac{284228}{15}\za{3}\za{7}-\frac{580888}{45}\za{1}\za{3}\za{7}\nn\\&-\frac{176624}{45}\za{3}^2\za{7}+\frac{122164}{45}\za{5}\za{7}-\frac{134176}{45}\za{3}\za{5}\za{7}\nn\\&-\frac{324968}{45}\za{7}^2+\frac{469616}{45}\za{1}\za{7}^2-\frac{18123}{50}\za{9}\nn\\&+\frac{34748}{5}\za{1}\za{9}-\frac{1343944}{75}\za{3}\za{9}+\frac{176624}{15}\za{1}\za{3}\za{9}\nn\\&+\frac{134176}{25}\za{3}^2\za{9}+\frac{185696}{15}\za{5}\za{9}-\frac{268352}{15}\za{1}\za{5}\za{9}\nn\\&
=-0.001365096861940014\cdots
\end{align}

\begin{align}
\zcorf{1}{2}{5}{6}{x}{z}{z}{x}&=
\frac{1}{240}-\frac{17}{20}\za{1}+\frac{872}{45}\za{3}-\frac{4688}{45}\za{1}\za{3}-\frac{24352}{45}\za{3}^2\nn\\&-\frac{27224}{135}\za{3}^3-\frac{77453}{450}\za{5}+\frac{107078}{45}\za{1}\za{5}+\frac{18952}{225}\za{3}\za{5}\nn\\&+\frac{54448}{45}\za{1}\za{3}\za{5}-\frac{82984}{225}\za{3}^2\za{5}-\frac{42242}{5}\za{5}^2\nn\\&+\frac{331936}{45}\za{1}\za{5}^2+\frac{12616}{9}\za{3}\za{5}^2+\frac{19168}{27}\za{5}^3+\frac{30961}{60}\za{7}\nn\\&-\frac{46109}{5}\za{1}\za{7}+\frac{284228}{15}\za{3}\za{7}-\frac{580888}{45}\za{1}\za{3}\za{7}\nn\\&-\frac{176624}{45}\za{3}^2\za{7}+\frac{122164}{45}\za{5}\za{7}-\frac{134176}{45}\za{3}\za{5}\za{7}\nn\\&-\frac{324968}{45}\za{7}^2+\frac{469616}{45}\za{1}\za{7}^2-\frac{18123}{50}\za{9}\nn\\&+\frac{34748}{5}\za{1}\za{9}-\frac{1343944}{75}\za{3}\za{9}+\frac{176624}{15}\za{1}\za{3}\za{9}\nn\\&+\frac{134176}{25}\za{3}^2\za{9}+\frac{185696}{15}\za{5}\za{9}-\frac{268352}{15}\za{1}\za{5}\za{9}\nn\\&
=0.001592580594206881\cdots
\end{align}

\begin{align}
\zcorf{1}{4}{5}{6}{x}{x}{z}{z}&=
\frac{1}{240}+\frac{1}{20}\za{1}-\frac{1093}{360}\za{3}+\frac{659}{45}\za{1}\za{3}+\frac{5621}{90}\za{3}^2\nn\\&+\frac{2704}{135}\za{3}^3+\frac{38411}{1800}\za{5}-\frac{2452}{9}\za{1}\za{5}-\frac{3857}{450}\za{3}\za{5}\nn\\&-\frac{5408}{45}\za{1}\za{3}\za{5}+\frac{8272}{225}\za{3}^2\za{5}+\frac{39289}{45}\za{5}^2\nn\\&-\frac{33088}{45}\za{1}\za{5}^2-\frac{1256}{9}\za{3}\za{5}^2-\frac{1904}{27}\za{5}^3-\frac{1589}{30}\za{7}\nn\\&+\frac{14581}{15}\za{1}\za{7}-\frac{17633}{9}\za{3}\za{7}+\frac{57904}{45}\za{1}\za{3}\za{7}\nn\\&+\frac{17584}{45}\za{3}^2\za{7}-\frac{1393}{5}\za{5}\za{7}+\frac{13328}{45}\za{3}\za{5}\za{7}\nn\\&+\frac{33124}{45}\za{7}^2-\frac{46648}{45}\za{1}\za{7}^2+\frac{1729}{50}\za{9}\nn\\&-714\za{1}\za{9}+\frac{137732}{75}\za{3}\za{9}-\frac{17584}{15}\za{1}\za{3}\za{9}\nn\\&-\frac{13328}{25}\za{3}^2\za{9}-\frac{18928}{15}\za{5}\za{9}+\frac{26656}{15}\za{1}\za{5}\za{9}\nn\\&
=0.009188091173609528\cdots
\end{align}

\begin{align}
\zcorf{1}{4}{5}{6}{x}{z}{x}{z}&=
\frac{1}{240}-\frac{11}{30}\za{1}+\frac{254}{45}\za{3}-\frac{144}{5}\za{1}\za{3}-\frac{6517}{45}\za{3}^2\nn\\&-\frac{7556}{135}\za{3}^3-\frac{18091}{450}\za{5}+\frac{5578}{9}\za{1}\za{5}+\frac{253}{75}\za{3}\za{5}\nn\\&+\frac{15112}{45}\za{1}\za{3}\za{5}-\frac{22928}{225}\za{3}^2\za{5}-\frac{105686}{45}\za{5}^2\nn\\&+\frac{91712}{45}\za{1}\za{5}^2+\frac{3484}{9}\za{3}\za{5}^2+\frac{5296}{27}\za{5}^3+\frac{9793}{90}\za{7}\nn\\&-\frac{106547}{45}\za{1}\za{7}+\frac{47236}{9}\za{3}\za{7}-\frac{160496}{45}\za{1}\za{3}\za{7}\nn\\&-\frac{48776}{45}\za{3}^2\za{7}+\frac{3682}{5}\za{5}\za{7}-\frac{37072}{45}\za{3}\za{5}\za{7}\nn\\&-\frac{91826}{45}\za{7}^2+\frac{129752}{45}\za{1}\za{7}^2-\frac{1848}{25}\za{9}\nn\\&+1778\za{1}\za{9}-\frac{371518}{75}\za{3}\za{9}+\frac{48776}{15}\za{1}\za{3}\za{9}\nn\\&+\frac{37072}{25}\za{3}^2\za{9}+\frac{52472}{15}\za{5}\za{9}-\frac{74144}{15}\za{1}\za{5}\za{9}\nn\\&
=-0.003158274321296133\cdots
\end{align}

\begin{align}
\zcorf{1}{4}{5}{6}{x}{z}{z}{x}&=
\frac{1}{240}-\frac{37}{60}\za{1}+\frac{2887}{360}\za{3}-\frac{1651}{45}\za{1}\za{3}-\frac{4813}{30}\za{3}^2\nn\\&-\frac{7556}{135}\za{3}^3-\frac{28313}{600}\za{5}+\frac{6124}{9}\za{1}\za{5}-\frac{7}{450}\za{3}\za{5}\nn\\&+\frac{15112}{45}\za{1}\za{3}\za{5}-\frac{22928}{225}\za{3}^2\za{5}-\frac{106436}{45}\za{5}^2\nn\\&+\frac{91712}{45}\za{1}\za{5}^2+\frac{3484}{9}\za{3}\za{5}^2+\frac{5296}{27}\za{5}^3+\frac{20461}{180}\za{7}\nn\\&-\frac{108892}{45}\za{1}\za{7}+\frac{47551}{9}\za{3}\za{7}-\frac{160496}{45}\za{1}\za{3}\za{7}\nn\\&-\frac{48776}{45}\za{3}^2\za{7}+\frac{3682}{5}\za{5}\za{7}-\frac{37072}{45}\za{3}\za{5}\za{7}\nn\\&-\frac{91826}{45}\za{7}^2+\frac{129752}{45}\za{1}\za{7}^2-\frac{1848}{25}\za{9}\nn\\&+1778\za{1}\za{9}-\frac{371518}{75}\za{3}\za{9}+\frac{48776}{15}\za{1}\za{3}\za{9}\nn\\&+\frac{37072}{25}\za{3}^2\za{9}+\frac{52472}{15}\za{5}\za{9}-\frac{74144}{15}\za{1}\za{5}\za{9}\nn\\&
=0.005680615367538651\cdots
\end{align}

\begin{align}
\zcorf{1}{3}{5}{6}{x}{x}{z}{z}&=
\frac{1}{240}+\frac{17}{60}\za{1}-\frac{2617}{360}\za{3}+\frac{1546}{45}\za{1}\za{3}+\frac{2935}{18}\za{3}^2\nn\\&+\frac{8008}{135}\za{3}^3+\frac{10811}{200}\za{5}-\frac{6430}{9}\za{1}\za{5}-\frac{3481}{150}\za{3}\za{5}\nn\\&-\frac{16016}{45}\za{1}\za{3}\za{5}+\frac{8216}{75}\za{3}^2\za{5}+\frac{111916}{45}\za{5}^2\nn\\&-\frac{32864}{15}\za{1}\za{5}^2-\frac{3752}{9}\za{3}\za{5}^2-\frac{5696}{27}\za{5}^3-\frac{13097}{90}\za{7}\nn\\&+\frac{121282}{45}\za{1}\za{7}-\frac{16730}{3}\za{3}\za{7}+\frac{57512}{15}\za{1}\za{3}\za{7}\nn\\&+\frac{52528}{45}\za{3}^2\za{7}-\frac{11921}{15}\za{5}\za{7}+\frac{39872}{45}\za{3}\za{5}\za{7}\nn\\&+\frac{95746}{45}\za{7}^2-\frac{139552}{45}\za{1}\za{7}^2+\frac{4921}{50}\za{9}\nn\\&-2016\za{1}\za{9}+\frac{394898}{75}\za{3}\za{9}-\frac{52528}{15}\za{1}\za{3}\za{9}\nn\\&-\frac{39872}{25}\za{3}^2\za{9}-\frac{54712}{15}\za{5}\za{9}+\frac{79744}{15}\za{1}\za{5}\za{9}\nn\\&
=-0.008335007472438759\cdots
\end{align}

\begin{align}
\zcorf{1}{3}{5}{6}{x}{z}{x}{z}&=
\frac{1}{240}-\frac{11}{20}\za{1}+\frac{1979}{180}\za{3}-\frac{833}{15}\za{1}\za{3}-\frac{4919}{18}\za{3}^2\nn\\&-\frac{13612}{135}\za{3}^3-\frac{75923}{900}\za{5}+\frac{10660}{9}\za{1}\za{5}+\frac{10547}{450}\za{3}\za{5}\nn\\&+\frac{27224}{45}\za{1}\za{3}\za{5}-\frac{41492}{225}\za{3}^2\za{5}-\frac{189064}{45}\za{5}^2\nn\\&+\frac{165968}{45}\za{1}\za{5}^2+\frac{6308}{9}\za{3}\za{5}^2+\frac{9584}{27}\za{5}^3+\frac{41321}{180}\za{7}\nn\\&-\frac{200018}{45}\za{1}\za{7}+\frac{28217}{3}\za{3}\za{7}-\frac{290444}{45}\za{1}\za{3}\za{7}\nn\\&-\frac{88312}{45}\za{3}^2\za{7}+\frac{59927}{45}\za{5}\za{7}-\frac{67088}{45}\za{3}\za{5}\za{7}\nn\\&-\frac{162484}{45}\za{7}^2+\frac{234808}{45}\za{1}\za{7}^2-\frac{3892}{25}\za{9}\nn\\&+3318\za{1}\za{9}-\frac{665042}{75}\za{3}\za{9}+\frac{88312}{15}\za{1}\za{3}\za{9}\nn\\&+\frac{67088}{25}\za{3}^2\za{9}+\frac{92848}{15}\za{5}\za{9}-\frac{134176}{15}\za{1}\za{5}\za{9}\nn\\&
=0.0008303788046606665\cdots
\end{align}

\begin{align}
\zcorf{1}{3}{5}{6}{x}{z}{z}{x}&=
\frac{1}{240}-\frac{43}{60}\za{1}+\frac{487}{40}\za{3}-\frac{2599}{45}\za{1}\za{3}-\frac{2491}{9}\za{3}^2\nn\\&-\frac{13612}{135}\za{3}^3-\frac{156821}{1800}\za{5}+\frac{3596}{3}\za{1}\za{5}+\frac{5161}{225}\za{3}\za{5}\nn\\&+\frac{27224}{45}\za{1}\za{3}\za{5}-\frac{41492}{225}\za{3}^2\za{5}-\frac{189214}{45}\za{5}^2\nn\\&+\frac{165968}{45}\za{1}\za{5}^2+\frac{6308}{9}\za{3}\za{5}^2+\frac{9584}{27}\za{5}^3+\frac{10409}{45}\za{7}\nn\\&-\frac{200543}{45}\za{1}\za{7}+\frac{28238}{3}\za{3}\za{7}-\frac{290444}{45}\za{1}\za{3}\za{7}\nn\\&-\frac{88312}{45}\za{3}^2\za{7}+\frac{59927}{45}\za{5}\za{7}-\frac{67088}{45}\za{3}\za{5}\za{7}\nn\\&-\frac{162484}{45}\za{7}^2+\frac{234808}{45}\za{1}\za{7}^2-\frac{3892}{25}\za{9}\nn\\&+3318\za{1}\za{9}-\frac{665042}{75}\za{3}\za{9}+\frac{88312}{15}\za{1}\za{3}\za{9}\nn\\&+\frac{67088}{25}\za{3}^2\za{9}+\frac{92848}{15}\za{5}\za{9}-\frac{134176}{15}\za{1}\za{5}\za{9}\nn\\&
=-0.001340720075108033\cdots
\end{align}

\begin{align}
\zcors{x}{x}{x}{x}{y}{y}&=
\frac{1}{2240}+\frac{11}{240}\za{1}-\frac{23}{16}\za{3}+\frac{382}{45}\za{1}\za{3}+\frac{7961}{180}\za{3}^2\nn\\&+\frac{148}{9}\za{3}^3+\frac{90583}{6300}\za{5}-\frac{5887}{30}\za{1}\za{5}-\frac{4061}{450}\za{3}\za{5}\nn\\&-\frac{296}{3}\za{1}\za{3}\za{5}+\frac{6716}{225}\za{3}^2\za{5}+\frac{430573}{630}\za{5}^2\nn\\&-\frac{26864}{45}\za{1}\za{5}^2-\frac{340}{3}\za{3}\za{5}^2-\frac{402}{7}\za{5}^3-\frac{32909}{720}\za{7}\nn\\&+\frac{69223}{90}\za{1}\za{7}-\frac{138161}{90}\za{3}\za{7}+\frac{47012}{45}\za{1}\za{3}\za{7}\nn\\&+\frac{952}{3}\za{3}^2\za{7}-\frac{39919}{180}\za{5}\za{7}+\frac{1206}{5}\za{3}\za{5}\za{7}\nn\\&+\frac{34811}{60}\za{7}^2-\frac{4221}{5}\za{1}\za{7}^2+\frac{6543}{200}\za{9}\nn\\&-\frac{2908}{5}\za{1}\za{9}+\frac{36322}{25}\za{3}\za{9}-952\za{1}\za{3}\za{9}\nn\\&-\frac{10854}{25}\za{3}^2\za{9}-\frac{4973}{5}\za{5}\za{9}+\frac{7236}{5}\za{1}\za{5}\za{9}\nn\\&
=-0.005996024922536831\cdots
\end{align}

\begin{align}
\zcors{x}{x}{x}{y}{x}{y}&=
\frac{1}{2240}-\frac{1}{240}\za{1}-\frac{643}{1440}\za{3}+\frac{179}{60}\za{1}\za{3}+\frac{1333}{90}\za{3}^2\nn\\&+\frac{223}{45}\za{3}^3+\frac{301507}{50400}\za{5}-\frac{6139}{90}\za{1}\za{5}-\frac{5221}{900}\za{3}\za{5}\nn\\&-\frac{446}{15}\za{1}\za{3}\za{5}+\frac{2018}{225}\za{3}^2\za{5}+\frac{251033}{1260}\za{5}^2\nn\\&-\frac{8072}{45}\za{1}\za{5}^2-34\za{3}\za{5}^2-\frac{1084}{63}\za{5}^3-\frac{14513}{720}\za{7}\nn\\&+\frac{11753}{45}\za{1}\za{7}-\frac{162149}{360}\za{3}\za{7}+\frac{14126}{45}\za{1}\za{3}\za{7}\nn\\&+\frac{476}{5}\za{3}^2\za{7}-\frac{24209}{360}\za{5}\za{7}+\frac{1084}{15}\za{3}\za{5}\za{7}\nn\\&+\frac{3241}{20}\za{7}^2-\frac{3794}{15}\za{1}\za{7}^2+\frac{731}{50}\za{9}\nn\\&-\frac{1959}{10}\za{1}\za{9}+\frac{21257}{50}\za{3}\za{9}-\frac{1428}{5}\za{1}\za{3}\za{9}\nn\\&-\frac{3252}{25}\za{3}^2\za{9}-\frac{1389}{5}\za{5}\za{9}+\frac{2168}{5}\za{1}\za{5}\za{9}\nn\\&
=0.001715129839332883\cdots
\end{align}

\begin{align}
\zcors{x}{x}{x}{y}{y}{x}&=
\frac{1}{2240}-\frac{7}{240}\za{1}-\frac{239}{1440}\za{3}+\frac{86}{45}\za{1}\za{3}+\frac{1123}{90}\za{3}^2\nn\\&+\frac{223}{45}\za{3}^3+\frac{82489}{16800}\za{5}-\frac{5309}{90}\za{1}\za{5}-\frac{2843}{450}\za{3}\za{5}\nn\\&-\frac{446}{15}\za{1}\za{3}\za{5}+\frac{2018}{225}\za{3}^2\za{5}+\frac{247883}{1260}\za{5}^2\nn\\&-\frac{8072}{45}\za{1}\za{5}^2-34\za{3}\za{5}^2-\frac{1084}{63}\za{5}^3-\frac{2785}{144}\za{7}\nn\\&+\frac{45563}{180}\za{1}\za{7}-\frac{160259}{360}\za{3}\za{7}+\frac{14126}{45}\za{1}\za{3}\za{7}\nn\\&+\frac{476}{5}\za{3}^2\za{7}-\frac{24209}{360}\za{5}\za{7}+\frac{1084}{15}\za{3}\za{5}\za{7}\nn\\&+\frac{3241}{20}\za{7}^2-\frac{3794}{15}\za{1}\za{7}^2+\frac{731}{50}\za{9}\nn\\&-\frac{1959}{10}\za{1}\za{9}+\frac{21257}{50}\za{3}\za{9}-\frac{1428}{5}\za{1}\za{3}\za{9}\nn\\&-\frac{3252}{25}\za{3}^2\za{9}-\frac{1389}{5}\za{5}\za{9}+\frac{2168}{5}\za{1}\za{5}\za{9}\nn\\&
=-0.002589603677721155\cdots
\end{align}

\begin{align}
\zcors{x}{x}{y}{x}{x}{y}&=
\frac{1}{2240}-\frac{11}{240}\za{1}+\frac{1537}{1440}\za{3}-\frac{421}{60}\za{1}\za{3}-\frac{317}{8}\za{3}^2\nn\\&-\frac{79}{5}\za{3}^3-\frac{623431}{50400}\za{5}+\frac{7949}{45}\za{1}\za{5}+\frac{15851}{1800}\za{3}\za{5}\nn\\&+\frac{474}{5}\za{1}\za{3}\za{5}-\frac{6404}{225}\za{3}^2\za{5}-\frac{808669}{1260}\za{5}^2\nn\\&+\frac{25616}{45}\za{1}\za{5}^2+108\za{3}\za{5}^2+\frac{3452}{63}\za{5}^3+\frac{15553}{360}\za{7}\nn\\&-\frac{64771}{90}\za{1}\za{7}+\frac{172993}{120}\za{3}\za{7}-\frac{44828}{45}\za{1}\za{3}\za{7}\nn\\&-\frac{1512}{5}\za{3}^2\za{7}+\frac{75247}{360}\za{5}\za{7}-\frac{3452}{15}\za{3}\za{5}\za{7}\nn\\&-\frac{32809}{60}\za{7}^2+\frac{12082}{15}\za{1}\za{7}^2-\frac{1593}{50}\za{9}\nn\\&+\frac{5503}{10}\za{1}\za{9}-\frac{68441}{50}\za{3}\za{9}+\frac{4536}{5}\za{1}\za{3}\za{9}\nn\\&+\frac{10356}{25}\za{3}^2\za{9}+\frac{4687}{5}\za{5}\za{9}-\frac{6904}{5}\za{1}\za{5}\za{9}\nn\\&
=-0.002089292530622660\cdots
\end{align}

\begin{align}
\zcors{x}{y}{x}{x}{x}{y}&=
\frac{1}{2240}-\frac{19}{240}\za{1}+\frac{95}{48}\za{3}-\frac{2087}{180}\za{1}\za{3}-\frac{22057}{360}\za{3}^2\nn\\&-\frac{3134}{135}\za{3}^3-\frac{12221}{630}\za{5}+\frac{2707}{10}\za{1}\za{5}+\frac{3971}{360}\za{3}\za{5}\nn\\&+\frac{6268}{45}\za{1}\za{3}\za{5}-\frac{1897}{45}\za{3}^2\za{5}-\frac{121259}{126}\za{5}^2\nn\\&+\frac{7588}{9}\za{1}\za{5}^2+\frac{1441}{9}\za{3}\za{5}^2+\frac{15340}{189}\za{5}^3+\frac{44329}{720}\za{7}\nn\\&-\frac{191611}{180}\za{1}\za{7}+\frac{194381}{90}\za{3}\za{7}-\frac{13279}{9}\za{1}\za{3}\za{7}\nn\\&-\frac{20174}{45}\za{3}^2\za{7}+\frac{2798}{9}\za{5}\za{7}-\frac{3068}{9}\za{3}\za{5}\za{7}\nn\\&-\frac{14777}{18}\za{7}^2+\frac{10738}{9}\za{1}\za{7}^2-\frac{1763}{40}\za{9}\nn\\&+\frac{4029}{5}\za{1}\za{9}-\frac{12269}{6}\za{3}\za{9}+\frac{20174}{15}\za{1}\za{3}\za{9}\nn\\&+\frac{3068}{5}\za{3}^2\za{9}+\frac{4222}{3}\za{5}\za{9}-\frac{6136}{3}\za{1}\za{5}\za{9}\nn\\&
=0.0008959339234083256\cdots
\end{align}

\begin{align}
\zcors{x}{x}{y}{x}{y}{x}&=
\frac{1}{2240}-\frac{13}{240}\za{1}+\frac{619}{480}\za{3}-\frac{229}{30}\za{1}\za{3}-\frac{14687}{360}\za{3}^2\nn\\&-\frac{79}{5}\za{3}^3-\frac{74279}{5600}\za{5}+\frac{2723}{15}\za{1}\za{5}+\frac{5197}{600}\za{3}\za{5}\nn\\&+\frac{474}{5}\za{1}\za{3}\za{5}-\frac{6404}{225}\za{3}^2\za{5}-\frac{810139}{1260}\za{5}^2\nn\\&+\frac{25616}{45}\za{1}\za{5}^2+108\za{3}\za{5}^2+\frac{3452}{63}\za{5}^3+\frac{2633}{60}\za{7}\nn\\&-\frac{8687}{12}\za{1}\za{7}+\frac{173287}{120}\za{3}\za{7}-\frac{44828}{45}\za{1}\za{3}\za{7}\nn\\&-\frac{1512}{5}\za{3}^2\za{7}+\frac{75247}{360}\za{5}\za{7}-\frac{3452}{15}\za{3}\za{5}\za{7}\nn\\&-\frac{32809}{60}\za{7}^2+\frac{12082}{15}\za{1}\za{7}^2-\frac{1593}{50}\za{9}\nn\\&+\frac{5503}{10}\za{1}\za{9}-\frac{68441}{50}\za{3}\za{9}+\frac{4536}{5}\za{1}\za{3}\za{9}\nn\\&+\frac{10356}{25}\za{3}^2\za{9}+\frac{4687}{5}\za{5}\za{9}-\frac{6904}{5}\za{1}\za{5}\za{9}\nn\\&
=0.001508610170462532\cdots
\end{align}

\begin{align}
\zcors{x}{y}{x}{z}{z}{y}&=
\frac{1}{6720}-\frac{1}{80}\za{1}+\frac{25}{288}\za{3}-\frac{16}{45}\za{1}\za{3}-\frac{121}{90}\za{3}^2\nn\\&-\frac{73}{135}\za{3}^3+\frac{1541}{50400}\za{5}+\frac{413}{90}\za{1}\za{5}-\frac{539}{450}\za{3}\za{5}\nn\\&+\frac{146}{45}\za{1}\za{3}\za{5}-\frac{211}{225}\za{3}^2\za{5}-\frac{32891}{1260}\za{5}^2\nn\\&+\frac{844}{45}\za{1}\za{5}^2+\frac{32}{9}\za{3}\za{5}^2+\frac{344}{189}\za{5}^3-\frac{137}{120}\za{7}\nn\\&-\frac{217}{15}\za{1}\za{7}+\frac{2578}{45}\za{3}\za{7}-\frac{1477}{45}\za{1}\za{3}\za{7}\nn\\&-\frac{448}{45}\za{3}^2\za{7}+\frac{107}{15}\za{5}\za{7}-\frac{344}{45}\za{3}\za{5}\za{7}\nn\\&-\frac{2219}{90}\za{7}^2+\frac{1204}{45}\za{1}\za{7}^2+\frac{207}{200}\za{9}\nn\\&+\frac{103}{10}\za{1}\za{9}-\frac{8017}{150}\za{3}\za{9}+\frac{448}{15}\za{1}\za{3}\za{9}\nn\\&+\frac{344}{25}\za{3}^2\za{9}+\frac{634}{15}\za{5}\za{9}-\frac{688}{15}\za{1}\za{5}\za{9}\nn\\&
=0.0007223711440396728\cdots
\end{align}

\begin{align}
\zcors{x}{x}{y}{z}{y}{z}&=
\frac{1}{6720}+\frac{1}{80}\za{1}-\frac{353}{720}\za{3}+\frac{569}{180}\za{1}\za{3}+\frac{327}{20}\za{3}^2\nn\\&+\frac{163}{27}\za{3}^3+\frac{26927}{5040}\za{5}-\frac{655}{9}\za{1}\za{5}-\frac{109}{30}\za{3}\za{5}\nn\\&-\frac{326}{9}\za{1}\za{3}\za{5}+\frac{488}{45}\za{3}^2\za{5}+\frac{63055}{252}\za{5}^2\nn\\&-\frac{1952}{9}\za{1}\za{5}^2-\frac{370}{9}\za{3}\za{5}^2-\frac{3940}{189}\za{5}^3-\frac{209}{12}\za{7}\nn\\&+\frac{4291}{15}\za{1}\za{7}-\frac{22497}{40}\za{3}\za{7}+\frac{3416}{9}\za{1}\za{3}\za{7}\nn\\&+\frac{1036}{9}\za{3}^2\za{7}-\frac{5869}{72}\za{5}\za{7}+\frac{788}{9}\za{3}\za{5}\za{7}\nn\\&+\frac{7609}{36}\za{7}^2-\frac{2758}{9}\za{1}\za{7}^2+\frac{251}{20}\za{9}\nn\\&-\frac{433}{2}\za{1}\za{9}+\frac{15961}{30}\za{3}\za{9}-\frac{1036}{3}\za{1}\za{3}\za{9}\nn\\&-\frac{788}{5}\za{3}^2\za{9}-\frac{1087}{3}\za{5}\za{9}+\frac{1576}{3}\za{1}\za{5}\za{9}\nn\\&
=0.001419281470187951\cdots
\end{align}

\begin{align}
\zcors{x}{x}{y}{z}{z}{y}&=
\frac{1}{6720}+\frac{1}{120}\za{1}-\frac{11}{24}\za{3}+\frac{571}{180}\za{1}\za{3}+\frac{733}{45}\za{3}^2\nn\\&+\frac{163}{27}\za{3}^3+\frac{26003}{5040}\za{5}-\frac{145}{2}\za{1}\za{5}-\frac{73}{20}\za{3}\za{5}\nn\\&-\frac{326}{9}\za{1}\za{3}\za{5}+\frac{488}{45}\za{3}^2\za{5}+\frac{63041}{252}\za{5}^2\nn\\&-\frac{1952}{9}\za{1}\za{5}^2-\frac{370}{9}\za{3}\za{5}^2-\frac{3940}{189}\za{5}^3-\frac{3107}{180}\za{7}\nn\\&+\frac{51443}{180}\za{1}\za{7}-\frac{67477}{120}\za{3}\za{7}+\frac{3416}{9}\za{1}\za{3}\za{7}\nn\\&+\frac{1036}{9}\za{3}^2\za{7}-\frac{5869}{72}\za{5}\za{7}+\frac{788}{9}\za{3}\za{5}\za{7}\nn\\&+\frac{7609}{36}\za{7}^2-\frac{2758}{9}\za{1}\za{7}^2+\frac{251}{20}\za{9}\nn\\&-\frac{433}{2}\za{1}\za{9}+\frac{15961}{30}\za{3}\za{9}-\frac{1036}{3}\za{1}\za{3}\za{9}\nn\\&-\frac{788}{5}\za{3}^2\za{9}-\frac{1087}{3}\za{5}\za{9}+\frac{1576}{3}\za{1}\za{5}\za{9}\nn\\&
=-0.002041037236271903\cdots
\end{align}

\begin{align}
\zcors{x}{y}{z}{x}{z}{y}&=
\frac{1}{6720}-\frac{7}{240}\za{1}+\frac{1169}{1440}\za{3}-\frac{941}{180}\za{1}\za{3}-\frac{3391}{120}\za{3}^2\nn\\&-\frac{1474}{135}\za{3}^3-\frac{150971}{16800}\za{5}+\frac{5662}{45}\za{1}\za{5}+\frac{1237}{200}\za{3}\za{5}\nn\\&+\frac{2948}{45}\za{1}\za{3}\za{5}-\frac{1474}{75}\za{3}^2\za{5}-\frac{281161}{630}\za{5}^2\nn\\&+\frac{5896}{15}\za{1}\za{5}^2+\frac{671}{9}\za{3}\za{5}^2+\frac{7148}{189}\za{5}^3+\frac{21893}{720}\za{7}\nn\\&-\frac{22673}{45}\za{1}\za{7}+\frac{5013}{5}\za{3}\za{7}-\frac{10318}{15}\za{1}\za{3}\za{7}\nn\\&-\frac{9394}{45}\za{3}^2\za{7}+\frac{2179}{15}\za{5}\za{7}-\frac{7148}{45}\za{3}\za{5}\za{7}\nn\\&-\frac{17059}{45}\za{7}^2+\frac{25018}{45}\za{1}\za{7}^2-\frac{4441}{200}\za{9}\nn\\&+\frac{1917}{5}\za{1}\za{9}
-\frac{71282}{75}\za{3}\za{9}+\frac{9394}{15}\za{1}\za{3}\za{9}\nn\\&+\frac{7148}{25}\za{3}^2\za{9}+\frac{9748}{15}\za{5}\za{9}-\frac{14296}{15}\za{1}\za{5}\za{9}\nn\\&
=-0.0001528594146694370\cdots
\end{align}

\begin{align}
\zcors{x}{y}{z}{z}{y}{x}&=
\frac{1}{6720}-\frac{1}{24}\za{1}+\frac{41}{36}\za{3}-\frac{109}{18}\za{1}\za{3}-\frac{571}{18}\za{3}^2\nn\\&-\frac{529}{45}\za{3}^3-\frac{134587}{12600}\za{5}+\frac{6328}{45}\za{1}\za{5}+\frac{2737}{450}\za{3}\za{5}\nn\\&+\frac{1058}{15}\za{1}\za{3}\za{5}-\frac{4852}{225}\za{3}^2\za{5}-\frac{617377}{1260}\za{5}^2\nn\\&+\frac{19408}{45}\za{1}\za{5}^2+82\za{3}\za{5}^2+\frac{872}{21}\za{5}^3+\frac{2599}{80}\za{7}\nn\\&-\frac{8197}{15}\za{1}\za{7}+\frac{10999}{10}\za{3}\za{7}-\frac{33964}{45}\za{1}\za{3}\za{7}\nn\\&-\frac{1148}{5}\za{3}^2\za{7}+\frac{7132}{45}\za{5}\za{7}-\frac{872}{5}\za{3}\za{5}\za{7}\nn\\&-\frac{2086}{5}\za{7}^2+\frac{3052}{5}\za{1}\za{7}^2-\frac{4581}{200}\za{9}\nn\\&+\frac{4121}{10}\za{1}\za{9}-\frac{26024}{25}\za{3}\za{9}+\frac{3444}{5}\za{1}\za{3}\za{9}\nn\\&+\frac{7848}{25}\za{3}^2\za{9}+\frac{3576}{5}\za{5}\za{9}-\frac{5232}{5}\za{1}\za{5}\za{9}\nn\\&
=-0.0003767104281403536\cdots
\end{align}
\end{appendix}

\end{document}